\documentclass[12pt, draftclsnofoot, onecolumn]{IEEEtran}

\usepackage{xcolor}
\usepackage[pagewise]{lineno}
\usepackage[utf8]{inputenc}
\usepackage{csquotes}
\usepackage[english]{babel}
\usepackage{array}
\usepackage{float}
\usepackage{enumitem}
\usepackage[normalem]{ulem}
\newcolumntype{P}[1]{>{\centering\arraybackslash}p{#1}}
\usepackage{multirow}
\usepackage{graphicx}
\usepackage{comment}
 \usepackage{caption}
\usepackage{subcaption}
\usepackage{url}

\usepackage{breakurl}
\usepackage{hyperref}
\hypersetup{breaklinks}
\usepackage{ulem}
\makeatletter
\usepackage{capt-of}

\begin{document}
%
\title{BeamSurfer: {Minimalist Beam Management of Mobile mm-Wave Devices}\footnote{This material is based upon work partially supported by US Department of Homeland Security under 70RSAT20CB0000017;
US Office of Naval Research under N00014-21-1-2385, N00014-18-1-2048;
US Army Research Office under W911NF-18-10331, W911NF-21-20064;
US Army Research Lab under W911NF-19-20243;
US National Science Foundation under OMA-2037890, Science $\&$ Technology Center Grant CCF-0939370, CNS-1719384.
The views expressed herein and conclusions contained in this document are those of the authors and should not be interpreted as representing the views or official policies, either expressed or implied, of the U.S. DHS, ONR, ARO, ARL, NSF,  or the United States Government.
The U.S. Government is authorized to reproduce and distribute reprints for Government purposes notwithstanding any copyright notation herein.}}
\author{\IEEEauthorblockN{Venkata Siva Santosh Ganji, Tzu-Hsiang Lin, Francisco A. Espinal, and P. R. Kumar, Fellow, IEEE}\\
\IEEEauthorblockA{Texas A\&M  University}}

\maketitle

\begin{abstract}
Management of narrow directional beams is critical for mm-wave communication systems. 
Translational or rotational motion of the user can cause misalignment of 
transmit and receive beams
with the base station losing track of the mobile. 
Reacquiring the user can take about one second in 5G New Radio systems
and significantly impair performance of applications, besides
being energy intensive.
It is therefore important to manage beams to continually maintain high received signal strength and prevent outages. 
It is also important to be able to recover from sudden but transient blockage
caused by a hand or face interposed in the Line-of-Sight (LoS) path.
This work presents a beam management protocol called BeamSurfer that {is targeted to the use case of users walking indoors near a base
station.
It is
designed to be minimalistic, employing only in-band information, and not requiring
 knowledge such as location or orientation of the mobile device or any additional sensors}.   
 Evaluations under pedestrian mobility show that {$96\%$ of the time} it employs beams that are within $3$ dB of what an omniscient Oracle would have chosen as the pair of optimal transmit-receive LoS beams.
It also recovers from transient LoS blockage by continuing to maintain control packet communication over a pre-selected reflected path that
preserves time-synchronization with the base station, which allows it to  rapidly recover to the LoS link as soon as it is unblocked.
\end{abstract}
\IEEEpeerreviewmaketitle

\section{Introduction}


Millimeter wave communication systems require narrow radio beams to combat high path loss, e.g., free space path loss of $84$  dB at $28$ GHz at $10$ m and penetration loss of at least $8$ dB from common building materials \cite{pathloss,_2018_study}. 

Ensuring that both receiving and transmitting beams are aligned ``towards each other"  while a receiver is mobile is one of the major challenges for mm-wave devices since misalignment results in weak received signal strength. 
A second major challenge 
is blockage of the LoS path that can happen without warning, e.g., by a user suddenly rotating and interposing her hand or face between transmitter and receiver.
It results in a rapid loss of signal strength of at least $10$ dB \cite{collonge_2004_influence}, shown in Fig. \ref{blockagemeasurement}c,
immediately resulting in
poor Signal to Noise Ratio (SNR) 
which leads to throughput degradation, increased packet re-transmissions, 
and possibly outage.
Whether due to misalignment or blockage, if communication between base station and user is disrupted, then the mobile is ``lost" to the base station, and has to be re-acquired as a new user, which incurs a delay of about one second, impairing applications.

The base station and mobile communicate with each other in what may be called  ``communication epochs", which are time intervals where the sender and the receiver are simultaneously using the appropriate transmit and receive beams, respectively, for each other, i.e., ``pointed" at each other. In each such communication epoch they can also make ``appointments" for a limited number of future communication epochs, so that they can continue to maintain communication with each other in the future. If communication between base station and mobile does not take place at the next few epochs, then the mobile is effectively ``lost" to the base station
because the
mobile loses time synchronization with the base station,
and also the base station and mobile do not know which transmit-receive beam pair will be effective for communicating after a delay.
The mobile will therefore need to be re-acquired all over again as a new user.
For acquiring new users,
mm-wave cellular base stations periodically transmit 
synchronization and reference signals.
In 5G New Radio technology, every $20$ ms, 
a synchronization signal block (SSB) 
is swept in up to $64$ directions \cite{_2017_nr2}. The acquisition delay can therefore be up to $1.28$ seconds for a mobile with single antenna array that is time unsynchronized with the base station, causing
a significant performance hit 
for
applications.
Re-acquisition also drains power from the mobile \cite{huang_2012_a,saha_2017_a}.
Avoiding \enquote{connection loss} is the primary goal of a beam management protocol.


{BeamSurfer targets the use case of users walking near 
an indoor base 
station. It employs two different mechanisms to avoid loss, for
user mobility which requires continuous adaptation, and
for blockage which is sudden and unpredictable.
It is a minimalist\footnote{The design philosophy of \enquote{less is more}, coined by Mies Van Der Rohe \cite{wiki:xxx}, is popular in the field of Architecture \cite{kim_2006}.
The architecture of mobile phones, especially the software and user interface, has embraced minimalist design from the
early 2010s.} solution in that it is a
totally in-band solution that uses the least amount of available information for minimizing connection loss, requiring no other knowledge such as location or orientation. Yet, it is effective, as we show experimentally that {$96\%$ of the time} it employs transmit and receive beams that are within $3$ dB of what an omniscient Oracle would have chosen.
}

\noindent
{\textbf{A. Outline of mobility adaptation.}}
We outline the adaptation mechanism for mobility; the details of the full protocol are provided in Section \ref{beamsurfersection}.
In each communication epoch both the sender and the receiver need to know what transmit or receive beams to use. For clarity of discussion, we suppose that the base station is the sender and the mobile is the receiver. 
The mobile need not reveal its receive beam to the base station. However, it needs to know that transmit beam the base station will use, which requires communication between them.

For the main lobe of an antenna radiation pattern, the directional gain drops steeply after half power, i.e., 3 dB loss.
  Its beamwidth is defined as twice the angular deviation at which there is a $3$ dB loss from maximum received signal strength.
  The experimental results presented in Sections
\ref{measurementsection}-B,C
confirm that the gain outside one beamwidth drops off significantly and is unpredictable. 
 As a user moves, BeamSurfer  first adapts the mobile's receiving beam to stay within $3$ dB of the best pair of transmit-receive beams, and when that does not suffice it provides feedback to the base station for it to adapt its beam to do so.
This ensures that it always utilizes a pair of beams that is within its beam width (half power points) where the gain is maximum.

As the user moves, the LoS path moves out of beam width, and 
we may call 
the duration after which the RSS drops below 3 dB from its maximum 
as the \textit{beam coherence time under mobility (BCT)} \cite{beamcoherencetime1}. 
It is how long a particular LoS beam is viable, similar to the notion of channel coherence time in wireless channels.
It also indicates how much time a mobile has to find a next beam to surf.
The primary factors affecting BCT are width of beam and user mobility,
as also confirmed by our 
experimental results for a 12 element phased array, presented in Section \ref{measurementsection}. 
The BCT is experimentally seen to be hundreds of milliseconds for human walking transversally to the LoS beam 10m from the base station.

Consider a system with azimuth beam steering. The number of non-overlapping beams covering a $120^{\circ}$ sector is $N:= \frac{120}{Beamwidth}$. Let us order them as $1,2, \ldots, N$. Suppose the current best beam is beam $k$. 
After mobility of hundreds of milliseconds, experimental results show that the next best beam is either $k-1$ or $k+1$, depending on whether the user continues moving/rotating in the same direction or changes direction.  
Upon detection of a $3$ dB loss, BeamSurfer employs an adaptation mechanism to determine which of $k-1$ or $k+1$ to switch to. Thereby, BeamSurfer ``surfs" the peaks of the transmit and receive beam lobes.
If neither $k-1$ or $k+1$ has a higher RSS, then the receiver asks the transmitter to try its neighboring beams. If there is steering in both azimuth and zenith directions, then there are eight neighbor beams to search over. The experimental evaluation of BeamSurfer in handling mobility is shown in Fig. \ref{beamsurferalignmentresults}.


While we have
focused on downlink,
if reciprocity holds, then, by reversing transmission and reception, the same pair of beams also works for uplink. 
In all our experiments, we found that beams are indeed reciprocal.
Even for potential non-reciprocal scenarios, the mobility adaptation can remain the same, 
except that separate beam-pairs will be needed for uplink and downlink.  

 The adaptation mechanism is completely in-band. It specifically does not require location of user, pose of antenna array, or any other side information  \cite{citation01,citation02,citation03,citation04,citation05}. Unlike many beam alignment algorithms \cite{aod1,aod2,aod3}, BeamSurfer does not require the LoS path's angle of arrival/departure estimates for beam alignment. Hence it can utilize  any 
 available beam patterns of any phased array \cite{phased_array}, and is applicable to any mm-wave technology.

\noindent
{\textbf{B. Outline of blockage recovery.}}
It is critical to avoid loss of the user for temporary blockages caused by an interposed hand or face between transmitter and receiver. 
Different mechanisms are needed to recover from such sudden transient blockages. 
The key to recovery is to preserve time-synchronization
during blockage.
The BeamSurfer protocol pre-scans and stores receive beam identities for a good reflected path.
Upon the sudden onset of blockage, BeamSurfer
 switches communication seamlessly to the pre-discovered good
reflected beam, which experiments show
has sufficient signal strength to  maintain connectivity and provide bidirectional communication with low-order modulation schemes for control packets.
This makes possible the critical functionality of preserving time-synchronization
during blockage.
It also makes it possible for the mobile to rendezvous back with the
base-station over an LoS path when the blockage ceases.

The solution presented is based on indoor experimental results
presented in Section \ref{measurementsection}-D
that the same transmitting beam
already being used by the base station for line of sight (LoS) communication with the user typically also
reaches
the user along a reflected path
with signal strength about $8$ to $10$ dB less than the LoS beam.
This reflected beam can be updated every $100$ ms and stored for use
in case blockage occurs.
The experimental results of Section \ref{measurementsection}-D show that it remains viable for that amount
of time even under mobility of the user. That is, the ``coherence time of reflected beams," defined more precisely in {Section \ref{measurementsection}-A}, is greater than $100$ ms. 

The mobile however needs to determine which is the best receive beam
for this reflected path.
This can be determined by scanning over all receive beams while the base station's transmit beam remains unchanged. 
Such opportunities arise because the base station periodically
sweeps over all transmit beams at
pre-announced times to acquire new users. 
 Knowing this schedule, \textit{and being already time-synchronized with the base station}, the mobile can scan over all its receive beams at those times when the station is using
the same transmit beam to acquire new users. Such dual usage of beams conserves valuable base station time. In our experiments we found that updating the NLOS beam only at instants when the transmit beam changes is sufficient to maintain a viable control channel.
A potential improvement is to also update to a new NLOS whenever the LoS receive beam changes, and to do so using compressive sensing methods \cite{Newref1,Newref6}.

While the signal strength of the reflected path is $8$ to $10$ dB lesser than for LoS, it is still
adequate for decoding control packets,
since those are transmitted by 
low order modulation schemes with polar or LDPC coding.
(In addition, it can also sustain data at a smaller rate up to $1$ Mbps).
The control plane messages can then be used to coordinate recovery from blockage by initiating a search for the best pair of transmit-receive beams over a reflected path.


Prior works \cite{reflect2, reflect3} backed by extensive measurement campaigns have already shown that mm-wave frequencies reflect from various surfaces in indoor environments and the reflected paths are usable for communication.  
The following two experimental results described in Section 
\ref{measurementsection}-D enable BeamSurfer's blockage recovery mechanism: \\
\noindent
1) The LoS transmit beam that is in current use 
typically also travels along a reflected path in the indoor environment, and
can be picked up by an appropriate receive beam with
signal strength within $10$ dB of that of the LoS beam.
Therefore, without switching the transmitting beam the reflected path can be reversed to allow the user to communicate with the base station after a blockage happens.
This is critical since the base station does not know that blockage has occurred, and so cannot change its transmit beam. It is only the mobile that knows that the blockage has occurred, and it needs to initiate blockage recovery. The signal strength of the reflected beam is sufficient to sustain such control plane communication. \\
\noindent
2) The receive beam that can pick up the reflected path remains viable for at least $100$ ms, so it can be determined by the mobile
ahead of blockage and stored.
Fig. \ref{blockagerecovery} shows the ability of BeamSurfer to react successfully to blockages in a typical office room environment.

 A well-founded beam management strategy needs to be of low complexity, have small realignment delay, provide good received signal strength/SNR, and maximally avoid disconnection from the network.
 BeamSurfer attains the above requirements
 by only employing
  data obtained during normal operation, {i.e., received signal strength}.
 

In Section \ref{measurementsection} we elaborate on the experiments. Then we present BeamSurfer's design 
in Section \ref{beamsurfersection}. 
The testbed is described in Section \ref{testbedsection}.
We present the
results of extensive mobility experiments in Section \ref{Implementation and Evaluation}.
\section{Related work} \label{Related work}
We now detail prior work on in/out of-band beam management and blockage recovery.
\textbf{A. In-band beam management.}
AgileLink \cite{beamalign1} makes use of the spatial sparsity of transmitted mm-wave signals
to align to a transmit beam in a logarithmic number of measurements.  BeamSurfer, in contrast, simply aligns by making measurements on neighbour beams. The mobility experiments and  evaluation of BeamSurfer show that employing neighbour beams ensures rapid alignment, resulting in beams that are nearly optimal with respect to the beams that an omniscient oracle would have chosen.
Beam-forecast \cite{beamalign2} and \cite{beamalign3} 
propose model-driven beam alignment. Beam-forecast first creates a geometric model extracting angles of arrival, phases and strengths of the paths, and then predicts the channel at a nearby location. It then uses another model to predict the best aligned TX-RX beam pairs. BeamSurfer does not require any computationally intensive model-based approaches, but nevertheless aligns within $3$ dB of oracle alignment.  
In \cite{beamalign4}, the authors propose a beam steering algorithm aims to reduce 
search space for re-alignment by using the history of aligned transmit and receive beam pairs. This approach works only if an associated beam pair already exists in memory. Furthermore this approach cannot be employed if either vertical or horizontal mobile orientation changes since the receive beam previously associated with a particular transmit beam will no longer be the right beam to use. 
BeamSurfer's approach, in contrast, is independent of the mobile's horizontal or vertical orientation and history of prior alignments, and can 
therefore handle beam alignment irrespective of
orientation of the mobile.
BeamSurfer is a lightweight beam management protocol not requiring any model parameter extraction \cite{beamalign2, beamalign3} or on-line determination of antenna element weights \cite{beamalign1}. Unlike those works discussed above that require storing several history-related parameters,
with consequent impact on memory and induced latency, BeamSurfer needs to store just one parameter -- the index of a good reflected beam for the current transmit beam.

{The problem of iteratively arriving at optimal antenna weights between a transmitter and a non-mobile receiver, when the channel is not changing, and when the two already have a functional two-way communication channel between them, is addressed in \cite{Newref1}, \cite{Newref6}, \cite{Newref2} and \cite{Newref5}.  These methods greatly reduce the number of measurements necessary, compared to exhaustive search, to determine optimal phase weights to align beams.} Using the spatial sparsity of the mm-wave channel, Swift-link \cite{Newref1} and FALP \cite{Newref6}.
%
investigate compressive sensing-based beam alignment. FALP and SwiftLink  propose  random spatial search to make a series of measurements and then identify the channel as well as the final antenna array phase weights 
for an aligned beam.
Swift-link also designs a training sequence.  In \cite{Newref2}, Hierarchical Beam Alignment (HBA) formulates beam alignment as a multi-arm bandit problem. 
In \cite{Newref5}, after the beam search space is optimized to a few dominant channel paths
a class of linear iterative search algorithms is proposed.

{The above works require accurate single antenna element measurements.  They may not be usable when the distance between base station and mobile  only permits communication over a pair of well aligned narrow transmit and receive beams, or when there is no separate control channel for feedback from receiver. In 
such restrictive situations, 
the problem becomes one of \textit{maintaining} alignment, once initial alignment has been achieved. This is the problem addressed by BeamSurfer through its ``surfing" process described in Section \ref{beamsurfersection}}. 
Rather than allowing for all phase matrices of the antenna array, BeamSurfer works in the space of beams for which phase matrix `codebooks" have been already defined through prior experimentation.
It effectively uses a priori defined ``neighbor beams" of the current beam to handle user mobility. The number of neighbor beams is eight for beams narrow in both azimuth and elevation, and two for planar beams only narrow in azimuth. The complexity of search for maintaining a well aligned beam is much smaller than the number of antenna elements considered in \cite{Newref1}, \cite{Newref6}, \cite{Newref2}, \cite{Newref5}.
For initial acquisition, the 3GPP standard using random access methods for transmit and receive beams. {In the 5G NR standard the base station sweeps transmit beams in $64$ directions periodically every $20$ milliseconds \cite{_2017_nr3}. The mobile scans over its receive beams to discover one or more of the base station's transmit beams. This procedure can take up to 1.28 seconds.}

The experimental studies in \cite{b1,b2,b3} highlight the negative impacts of body blockage. Beamspy \cite{beamspy} uses spatially correlated beams and a
model driven framework to overcome human blockage.
It constructs a path skeleton with fine grain measurements using nonlinear least squares, and uses the channel impulse response to predict unblocked paths. Beam switching \cite{blockage1}
also relies on an environment model to avoid blockage. The fundamental issue with model based approaches is that unlike user mobility, hand and body blockage is sudden and unpredictable.
A hand grip-aware blockage recovery scheme is proposed in  
\cite{blockage2}. 
In contrast, BeamSurfer harvests reflected beams without requiring a model, and handles blockage at run-time. By extracting path parameters of mm-wave channel and making simultaneous measurements using {four antenna arrays placed at the four corners of the mobile, a Fast Antenna and Beam Switching method is proposed in \cite{Newref7}
to address hand blockage. Using channel parameters estimated for the active antenna array, the method attempts to predict the channel for all the other inactive arrays, with the goal of activating the antenna array with better channel estimates.} This requires additional on-chip hardware. Moroever it cannot predict the channel state for the other inactive arrays without knowledge of their locations and mobile orientation.  


\noindent
\textbf{B. Out-of-band beam management.}
 Sensor assisted beam alignment is studied in \cite{citation05,sensor_issues,sideinfo2} employing information from motion sensors such as accelerometers, gyroscopes  and magnetometers. The approach employed in \cite{sideinfo1} needs angle of arrival to handle beam misalignment.
Pia \cite{citation02} uses pose of the array to correct beam misalignment. Light sensors \cite{citation01} and side channel information \cite{citation03, citation04, sensor3} have also been suggested to manage antenna beams.

Through extensive experimentation, \cite{sensor_issues} has studied the major challenges involved in using sensors available on the mobile phones, such as accelerometers, gyroscopes and magnetometers, for beam management.  It observed that speed and accuracy of sensor information is critical for aligning beams.  Sensor error required to align beams of $30^\circ$ beam width for small rotational movements is under $1.2^\circ$.  The authors point out that the currently available maximum sampling rates on high end mobile phones,  100 Hz for the  accelerator and 200 Hz for the gyroscope, are insufficient to predict small and quick angular movements that happen while mobile user is playing games.  Moreover, even these sampling rates are only available in the high power usage mode of sensors which is not ideal for mobile devices.
  Another approach studied in \cite{Newref4} is to extract spatial information from the omnidirectional sub-$6$ GHz link for beam selection. However, the channel behavior in the mm-wave bands for LoS channels is very different.
In contrast to all the above, BeamSurfer is a completely in-band solution that does not require any prior information about the local environment or any additional sensors.  
\section{The Goals and Challenges of Beam management}
Unlike omni-directional communication systems where interference can impact the Signal-to-Interference plus Noise Ratio (SINR), mm-wave systems using narrow directional beams are less susceptible to signal contamination from neighbor devices. Therefore, maintaining good signal strength is sufficient. Maintaining high signal strength, and
avoiding
user loss 
are the primary goals of this paper.
%
%
We explore a totally in-band solution,
based purely on information
normally acquired over the radio domain,
and not requiring any
out-of-band sensors.

The following are our key experimental findings:
\begin{enumerate}[topsep=0pt,itemsep=-1ex,partopsep=1ex,parsep=1ex]
\item For a given mobility scenario, there is a certain duration,
    which we call the Beam Coherence Time (BCT), for which a prior aligned  Transmit-Receive beam pair remains usable.  This BCT is typically greater than $100$ ms 
    even when the user rotates and interposes her face to block the LoS beams. Tables  \ref{table:1} $\&$ \ref{rotation_table}  present BCT for various motion patterns.
    BeamSurfer adapts  measurement intervals to this BCT to minimize probing or alignment attempts.
     \item When receive beam $k$ of the mobile gets misaligned, the adjacent receive beams $k-1$ and $k+1$ are a good starting point for re-aligning the beam. The heatmaps in Figs. \ref{fig:heatmaptrans} and \ref{rotationmeasure} presenting the signal strengths measured at the mobile on all receive beams 
     reveal that adjacent beams are indeed good choices to restore received signal strength.
     \item When receive beam adjustment by the mobile is not enough to compensate for mobility, the
     transmit beam needs to be changed. If the transmit beam
     $n$ is misaligned, the neighboring transmit beams $n-1$ and $n+1$ pointed in the adjacent
     directions are good starting points for re-aligning the transmit beam. The snapshots of experiments during BeamSurfer beam adaptation in Fig. \ref{beamsurferalignmentresults} identify occasions of transmit and receive beam adaptation.

      \item It is fine to choose a slightly misaligned beam first and refine it to reach a near-optimal pair of transmit-receive beams, rather than exploring afresh for the best beam pair. 
    \item The transmit beam used for LoS communication typically also has a reflected path with $8$ to $10$ dB lesser signal strength in indoor environments. The receive beam must be changed to pick up this reflected path. Currently this is done by exhaustive receive beam scanning, but compressive sensing methods \cite{Newref1,Newref6} are potentially useful here. This is an adaptation that can be performed by the mobile without
    requiring any coordination with the base station.   
    This makes it possible to recover from blockage, which can be initiated by the mobile. {Signal strength measured on all beams and presented as heatmap in Fig. \ref{blockagemeasurement}c during our LoS blockage experiments demonstrate the existence of usable NLoS paths.}
    \item  
    Blockage recovery by switching to a reflected beam also
    needs to be adaptive since
    the wireless environment is dynamic and the scattering changes rapidly  with motion.
    Because the BCT of reflected beams is greater than $100$ ms,
    it is adequate for the mobile to scan for a reflected beam every $100$ ms or every time the
    base station switches to a new LoS transmit beam, and store it for future usage.
    The receive beam so found for a transmit beam within the past $100$ ms serves as a fallback when blockage happens.
    \item The reflected beam's received signal strength that is about $8$ to $10$ dB lower than the
    LoS beam prior to blockage, is still adequate to maintain
    connectivity.
    \item An already found reflected path can be subsequently used to recover from a blockage caused outage since control packets can be bi-directionally communicated over the reflected path. 
    \item 
    Recovery from blockage to a better transmit-receive beam pair can therefore be performed seamlessly as long as control packets
    can be exchanged between mobile and base station.
\end{enumerate}

Based on this experimental evidence, we compose a near-optimal solution for beam management,
applicable to any mm-wave network, whether WiGig or 5G with directional beams.
We specifically describe how it can be implemented in 5G mm-wave standards.
\section{Measurements under mobility} \label{measurementsection}

\begin{figure}[h]
\centering
  \includegraphics[width=0.6\linewidth,scale=.3]{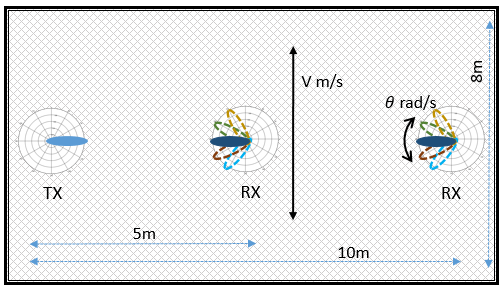}
  \captionof{figure}{Layout for experiments}
  \label{layout}
\end{figure}

To study how received signal strength for a directional beam changes with user motion, we performed repeated mobility experiments in both indoor and outdoor environments. The indoor setup is shown in Fig. \ref{layout}. The Testbed used for our experiments is detailed in Section \ref{testbedsection}.  

\noindent
\textbf{A. Beam Coherence Time.}

We determine BCT measured at the receiver for translational as well as rotational user motions in both indoor and outdoor environments, as described next.
\begin{figure*}[t]

	\begin{subfigure}[t]{.3\linewidth}
		\centering
		\includegraphics[width=2.2in]{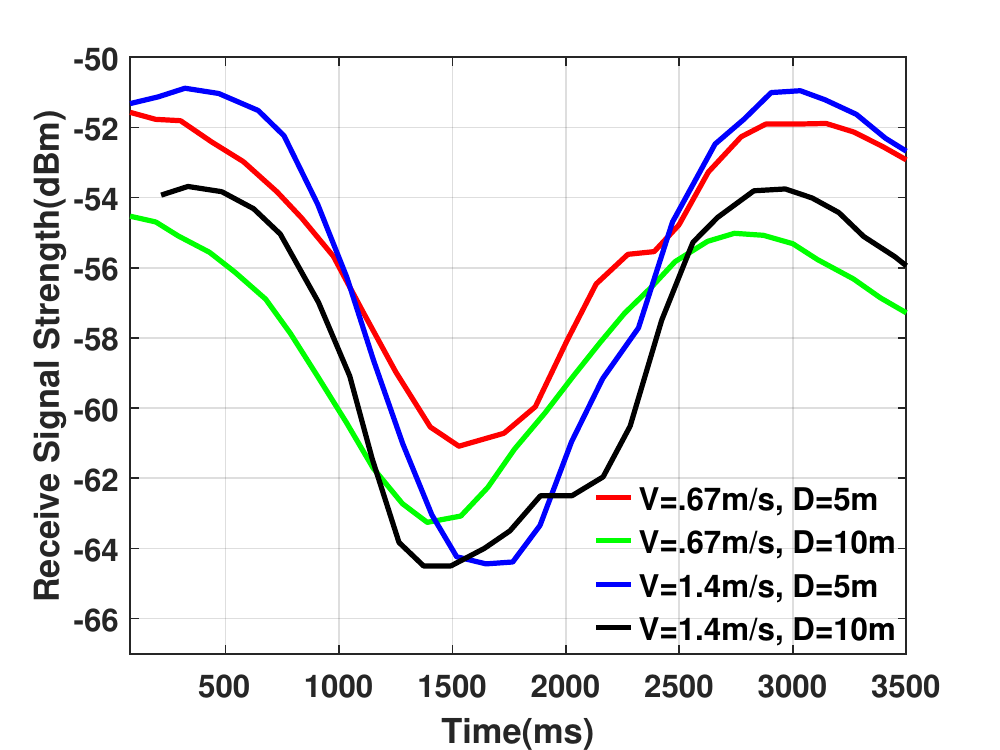}
		\caption{Indoor - Wide Beam}\label{fig:2}
	\end{subfigure}
		\quad
	\begin{subfigure}[t]{.3\linewidth}
		\centering
		\includegraphics[width=2.2in]{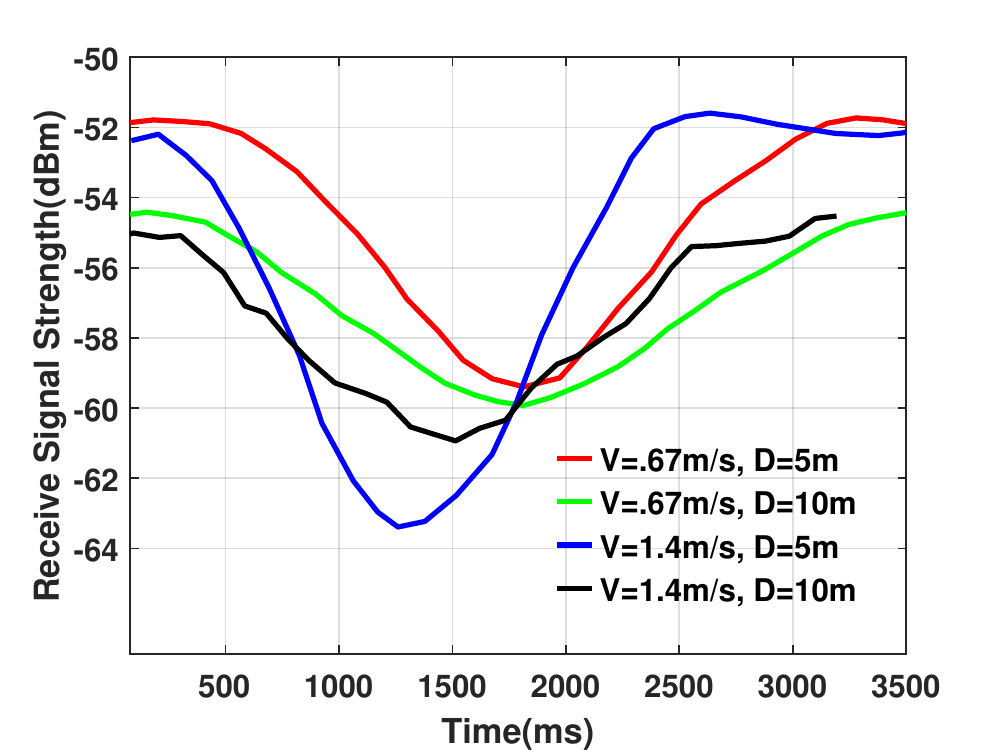}
		\caption{Outdoor- Narrow Beam}\label{fig:3}
	\end{subfigure}
		\quad
	\begin{subfigure}[t]{.3\linewidth}
		\centering
		\includegraphics[scale=.55]{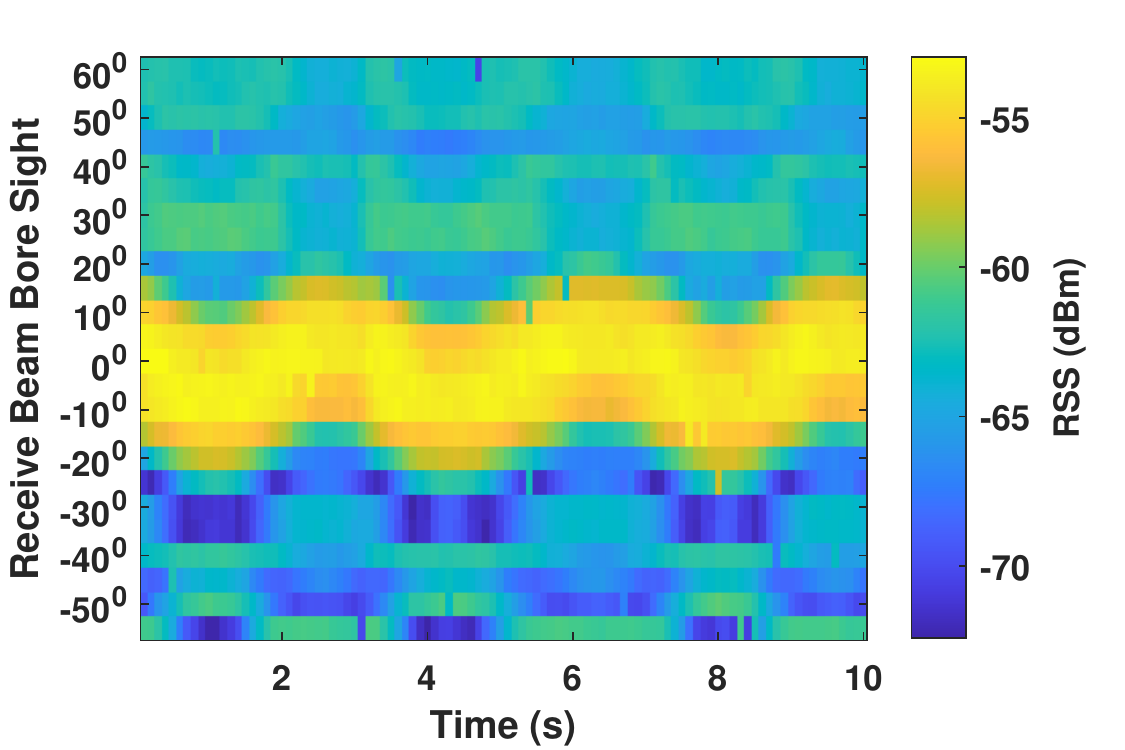}
		\caption{Heatmap}\label{fig:heatmaptrans}
	\end{subfigure}
	\caption{Impact on RSS: Lateral Translational Motion}\label{bct_walk}
\end{figure*}

\noindent
\textbf{B. Lateral Translational Motion.}
We positioned the transmitter at a fixed location as shown in Fig. \ref{layout}, with the beam always pointing at $0^{\circ}$ azimuth. The receiver node placed on a cart is moved $2$ to $4$ m perpendicular to the transmit beam direction at different speeds. 
Received Signal Strength and SNR are recorded with the receiver node  $5$ m and $10$ m away from the transmitter node. Each experiment is repeated $15$ times at $20$ different  receiver locations. The receiver beam is changed every $2$ ms with boresight angles sweeping from $60^{\circ}$ to $-60^{\circ}$. {The BCT is determined from the graphs in Fig. \ref{bct_walk}a and 2b as the time taken for the Received Signal Strength to decrease by 3 dB from its maximum.} The mean BCT is calculated for all the scenarios using both narrow {($20^{\circ}$)} and wide {($30^{\circ}$)} receive beams. 
{ The standard deviation of the RSS within the BCT is small, ranging from 0.64 dB for V= 0.67 m/s motion at 10 m to 1.12 dB for V=1.4 m/s motion at 5 m}.
Similar experiments are performed in the outdoor environments.

\begin{table}[h]
\caption{Mean BCT: Translational Motion}\label{table:1}
\small 
	\centering
	\begin{subtable}[t]{1.5in}
		\centering
		\begin{tabular}{ |p{.75cm}|p{1.1cm}|p{.75cm}| }
\hline
 \hline
 Speed (m/s)  & Distance (m) & BCT (ms)\\ 
 \hline\hline
  0.67 & 5 & 643  \\ 
 0.67 & 10 & 861  \\ 
 1.4 & 5 & 476  \\ 
 1.4 & 10 & 802  \\ 
 \hline
\hline
\end{tabular}
		\caption{Indoor: narrow beam}\label{table:1a}
	\end{subtable}
	\quad
	\begin{subtable}[t]{1.5in}
		\centering
		\begin{tabular}{ |p{.75cm}|p{1.1cm}|p{.75cm}| }
\hline
 \hline
 Speed (m/s)  & Distance (m) & BCT (ms)\\ 
 \hline\hline
  0.67 & 5 & 683  \\ 
 0.67 & 10 & 921  \\ 
 1.4 & 5 & 576  \\ 
 1.4 & 10 & 882  \\ 
 \hline
\hline
\end{tabular}
		\caption{Indoor: wide beam}\label{table:1b}
	\end{subtable}
		\begin{subtable}[t]{1.5in}
		\centering
		\begin{tabular}{ |p{.75cm}|p{1.1cm}|p{.75cm}| }
\hline
 \hline
 Speed (m/s)  & Distance (m) & BCT (ms)\\ 
 \hline\hline
  0.67 & 5 & 655  \\ 
 0.67 & 10 & 880  \\ 
 1.4 & 5 & 491  \\ 
 1.4 & 10 & 819  \\ 
 \hline
\hline
\end{tabular}
		\caption{Outdoor: narrow beam}\label{table:1a}
	\end{subtable}
	\quad
	\begin{subtable}[t]{1.5in}
		\centering
		\begin{tabular}{ |p{.75cm}|p{1.1cm}|p{.75cm}| }
\hline
 \hline
 Speed (m/s)  & Distance (m) & BCT (ms)\\ 
 \hline\hline
  0.67 & 5 & 701  \\ 
 0.67 & 10 & 933  \\ 
 1.4 & 5 & 589  \\ 
 1.4 & 10 & 896 \\ 
 \hline
\hline
\end{tabular}
		\caption{Outdoor: wide beam}\label{table:1b}
	\end{subtable}
	
\end{table}
 The BCTs for different TX-RX distances and speeds of lateral motion are presented in Table $1$.
 The variation of the received signal strength with motion is shown in Fig. $2$. Fig. $2$c shows a heatmap of the received signal as a mobile with a phased array is moved back and forth laterally three times
 at $1.4$ m/s, the nominal walking speed. For each of the receive beam directions from -$50^{\circ}$ to $+60^{\circ}$, it presents the received signal strength coded as a color from yellow (high $-45$ dBm) to dark blue (low $-75$ dBm). 

Three important conclusions may be drawn from the experiments.\\
\textbf{\textit{Observation 1.}} As shown in Table $1$ and Fig. $2$, the BCT is several hundreds of milliseconds for walking speeds for both indoor and outdoor environments. Therefore there is ample time to adapt and switch transmit-receive beam pairs for users moving at typical walking speeds.
\\
\textbf{\textit{Observation 2.}} Fig. \ref{fig:heatmaptrans} shows that there is smooth fall of signal strength as beam direction is changed
on both directions of the best beam. Importantly, a neighboring beam of maximal signal strength is second best, and, as the user moves, it becomes the best beam. This can be seen in Fig. 
\ref{fig:heatmaptrans} where one can track the best beam as the user moves back and forth. 
This considerably simplifies the beam management process. When the beam degrades, only the two neighboring beams are needed to determine the best configuration.
The best beam, $k$, can be used for a few hundred milliseconds until the signal strength degrades by $3$ dB. Then, the user can switch to either beam $k-1$ or $k+1$. The procedure is repeated until it finds the  optimal beam. This is readily doable using  mechanisms in $5$G NR standards, as there are quite a number of measurement opportunities \cite{_2017_nr2} to decide the neighbor beam even for full 3-D space.

Our results are based on experimental data collected over 2 months at different times of day, with 100 plus hours of mobility data in complex indoor environments: a lab with several metallic reflectors on walls, corridors and class rooms, as well as outdoor environments like pavements, large open places surrounded by buildings on all sides, and on public lawns. 

More generally, just as different scenarios can have different channel models or delay spread profiles, different mobility scenarios and environments can also have different BCTs. One may empirically calculate BCTs for such scenarios to tune parameters of the protocol. 

\begin{figure}
\centering
\begin{minipage}{0.35\textwidth}
\centering
\includegraphics[width=1.1\linewidth,height=1.8in]{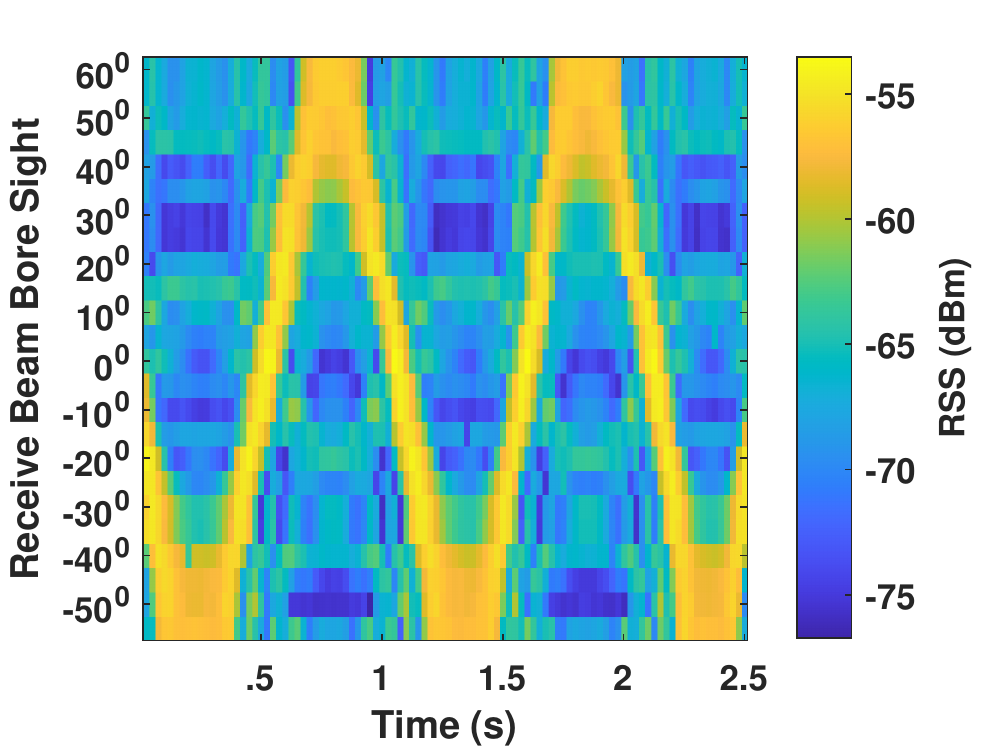} 
 \caption{Rotational Motion: Heatmap}
\label{rotationmeasure}
\end{minipage}
\begin{minipage}{0.55\textwidth}
\centering
\captionsetup{type=table} 
\small 
\caption{Mean BCT: Rotational Motion}\label{rotation_table}
\begin{center}
	\begin{subtable}[t]{1.5in}
		\centering
		\begin{tabular}{ |p{.75cm}|p{1.1cm}|p{.75cm}| }
\hline
 \hline
 Speed (rad/s)  & Distance (m) & BCT (ms)\\ 
 \hline\hline
  2$\pi$/9 & 5 & 284  \\ 
 $\pi$/3 & 5 & 200  \\ 
  2$\pi$/3 & 5 & 141  \\ 
 4$\pi$/3 & 5 & 101  \\ 

 \hline
\hline
\end{tabular}
		\caption{Narrow beam}\label{table:1a}
	\end{subtable}
	\quad
	\begin{subtable}[t]{1.5in}
		\centering
		\begin{tabular}{ |p{.75cm}|p{1.1cm}|p{.75cm}| }
\hline
 \hline
 Speed (rad/s)  & Distance (m) & BCT (ms)\\ 
 \hline\hline
  2$\pi$/9 & 5 & 300  \\ 
 $\pi$/3 & 5 & 219  \\ 
 2$\pi$/3 & 5 & 160  \\ 
 4$\pi$/3 & 5 & 124  \\ 
 \hline
\hline
\end{tabular}
		\caption{Wide beam}\label{table:1b}
	\end{subtable}
\end{center}

\end{minipage}
\end{figure}

\noindent
{\textbf{C. Rotational Motion.}}
Similar to lateral translational motion, we repeated rotation experiments within the same layout, as shown in Fig. \ref{layout}. The receiver phase array is rotated $120^{\circ}$ at multiple angular speeds. After $15$ trials at $20$ different receiver positions, we determined the mean BCT. Table \ref{rotation_table} lists the variables of the rotational motion experiment. Fig. \ref{rotationmeasure} shows the variation of received signal strength with rotational motion. We present BCT in the indoor environment only as we did not observe much difference outdoors. The BCT is shorter than that for lateral motion, but still greater than $100$ ms.

Fig. \ref{rotationmeasure} presents the heatmap for an angular speed of $120^{\circ}$/s. It exhibits the smoothness of received signal strength with beam direction at all times, as well as continuous fall off when one moves away in either direction from the best beam at any time.

For both motion patterns, translational and rotational, these observations show that simple beam adaptation mechanisms suffice by employing a beam till its received signal strength degrades by $3$ dB, and then switching to the better of its clockwise or counterclockwise beams, even under the more stressful rotational motion of mobiles.

\captionsetup[subfigure]{labelformat=empty}
  \begin{figure*}[t]
	\centering
	\begin{subfigure}[t]{.3\linewidth}
		\centering
		\includegraphics[width=2.2in,keepaspectratio]{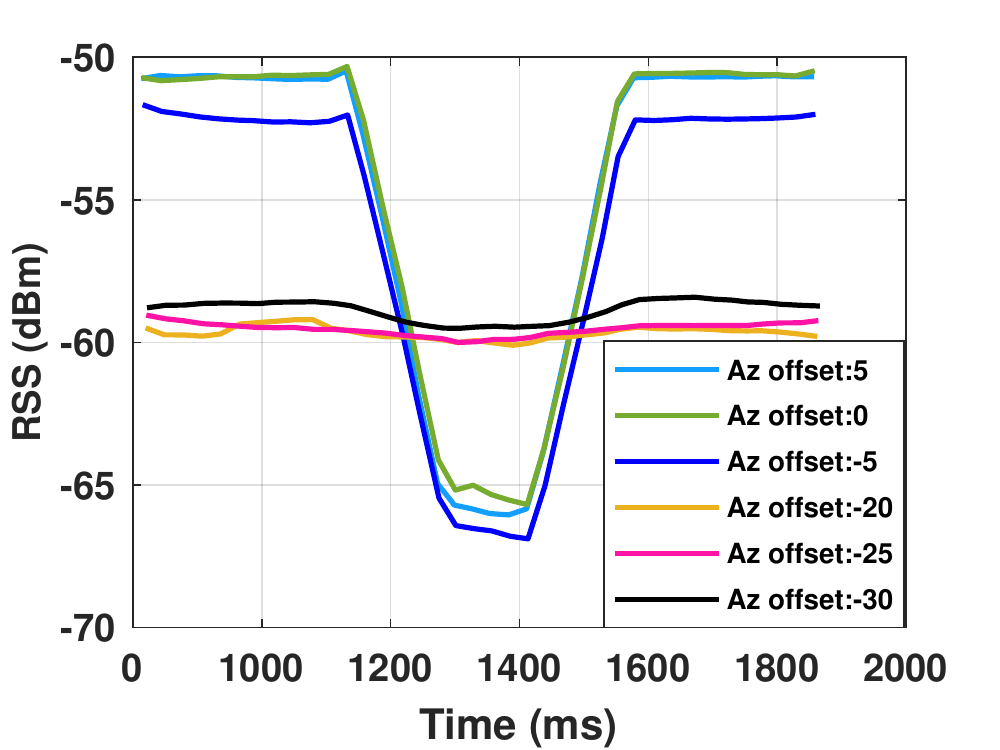}
		\caption{(a)}	
		\label{fig:b1}	
	\end{subfigure}
	\quad
  \begin{subfigure}[t]{.3\linewidth}
		\centering
		\includegraphics[width=2.2in,keepaspectratio]{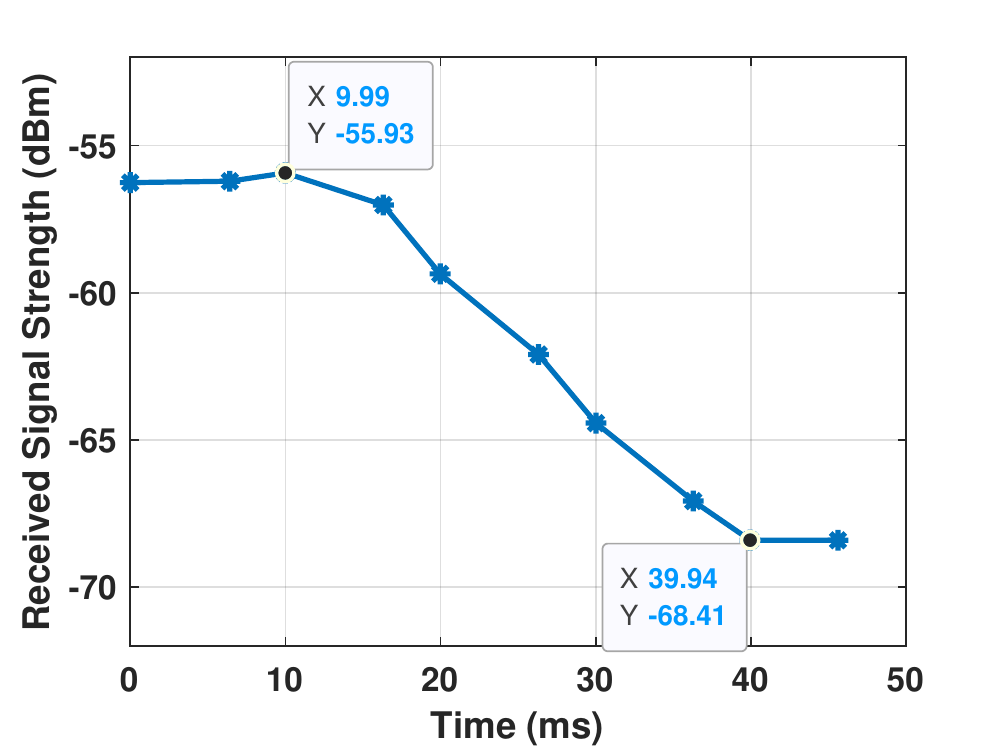}
		\caption{(b)}	
		\label{fig:b2}
	\end{subfigure}
	\quad
	\begin{subfigure}[t]{.3\linewidth}
		\centering
		\includegraphics[width=2.2in,keepaspectratio]{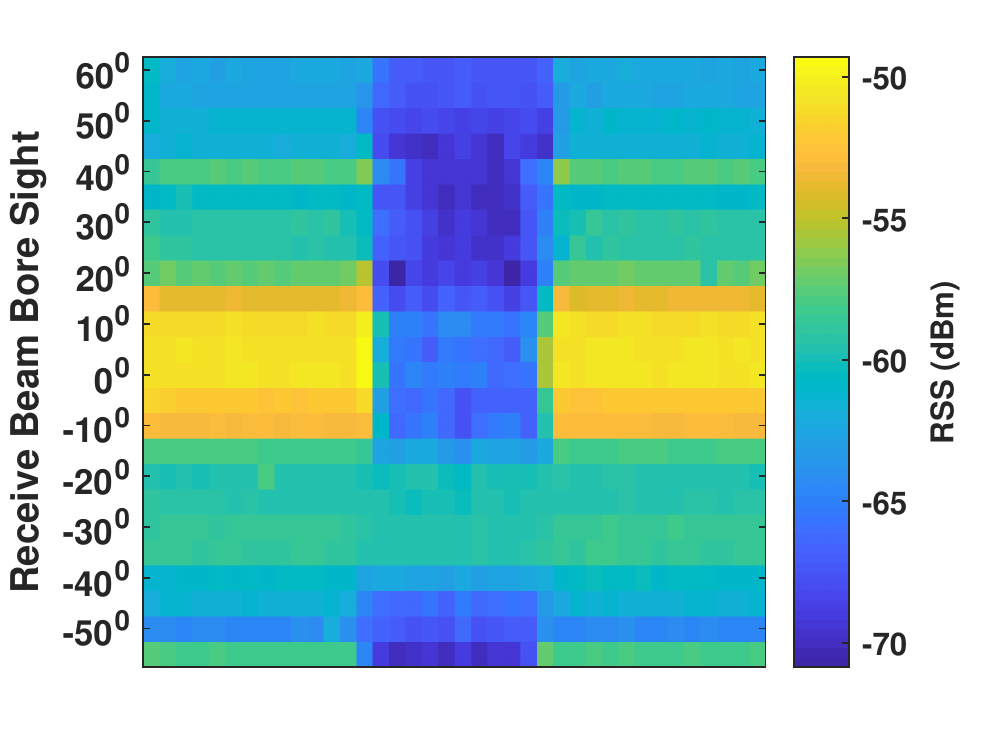}
		\caption{(c)}	
				\label{blockage}

	\end{subfigure}
		\caption{Impact of human body blockage on RSS. (a) LoS and NLoS (reflected) beams (b) Rapid hand movements (c) Heatmap.}
		\label{blockagemeasurement}
\end{figure*}

\noindent
{\textbf{D. Human Body Blockage.}}
Hand and human body movements can block mm-Wave signals, resulting in the RSS at the handheld mobiles dropping by several dBs, causing an outage. (Human fingers generally do not cause link outage. We observed that it is difficult to block all the elements of an antenna array, and even a couple of exposed antenna elements in the array can provide decent received signal strength, around $-58$ dBm at full bandwidth).

To measure impact of body blockage, we kept transmitter and receiver  $5$ m apart in Fig. \ref{layout}. The setup is placed $2$ m away from the wall, which reflects the transmit beam.  Blocking then occurrs by a person's hands, head, or body interposing between receiver and transmitter.

\textbf{\textit{Observation 4.}} Fig. \ref{blockagemeasurement}a shows the results of an experiment where blockage of LoS happens at around a second.
The RSS for the fixed transmit beam used by the base station is shown
during the course of the experiment for various receive beams.
In the beginning, there is no blockage of LoS. The transmit beam is received by an optimally aligned receive beam. The received signal strength remains  at 
$-51$ dBm for one second, at which time blockage happens.
{ Fig. \ref{blockagemeasurement}b} shows the result of hand blockage of the LoS by grabbing a mobile near the location of the antenna array \cite{blockage2}. It take place rapidly over approximately $30$ ms.
After blockage, the best receiving beam could still pick up the transmit beam through
a reflected path,
but at a loss of $8$ dB compared to
the prior LoS reception.
So, the same transmit beam can continue to be used at $8$ dB lower strength, but the mobile needs to adjust its receive beam
to pick up the reflected transmission.

\textbf{\textit{Observation 5.}}
Since the received signal strength of the reflected beam {from various surfaces like walls, tables, doors, etc.,} is only $8$ to $10$ dB lower than the direct LoS beam,
the control packets that are sent by the receiver using lower modulation and
coding (MCS) schemes can be decoded at the transmitter. 
This makes it possible to maintain connectivity of the link even after
blockage for control packets.
Upon receiving these control packets, transmitter can choose to adapt its beams to overcome blockage.
Importantly, clock synchronization can be maintained during blockage of the LoS.

\textbf{\textit{Observation 6.}}
 Fig. \ref{blockagemeasurement}c shows the heatmap of the signal strength of several receive beams during the course of the experiment. The y-axis shows the signal strength of various beams in a color coded fashion.
The occlusion of LoS receive beams of boresights $-5^\circ$ to $5^\circ$  is clearly seen, with the color of the RSS dropping from that corresponding to a high value of $-51$ dBm to a reduced RSS of $-66$ dBm. 
However, one may observe that the receive beams along the boresight angles $-25^\circ$ to $-40^\circ$ 
 maintain constant received signal strength throughout the experiment, unchanged by blockage.
This is because they are throughout received along reflected paths that are unaffected by
blockage.

\textbf{\textit{Observation 7.}} Another observation is that the receive beam chosen after blockage remained unaffected, and is usable throughout for the
same BCT as LoS beams.
This can be seen from Fig. \ref{blockagemeasurement}a
as the best receive beam after blockage remains
stable for the same BCT as LoS beams before blockage.
This shows that the best receive beam to pick up the reflected transmit beam can be chosen
$100$ ms ahead of time.
This allows the protocol to prepare for blockage by scanning and updating for a
suitable receive beam to pick up the reflected
transmit beam every $100$ ms.

The above observations are used to compose the blockage recovery protocol
of BeamSurfer by decoupling
the overall problem of handling blockage into two sub problems: \\
(i) \textbf{Being ready in case blockage happens}: If the user tries to determine a good receive beam \emph{after} blockage happens, then it (a) wastes 
the valuable data transmission time that was allocated by the base station to the user, and (b) wastes the base station's time. Instead it is better to determine a good receive beam
before blockage without wasting either the base station's time or the user's data transmission time. This can be done by sampling the transmit beam, and in 5G NR there are frequent opportunities to sample
for a mobile that is already synchronized with the base station. 
The user simply samples all the receive beams and stores the best receive beam
to pick up a reflected path for possible later use. 
The sampling is done after a new transmit beam is used or every $100$ ms, whichever is earlier, as our BCT experiments suggested $100$ ms is a good measurement interval to handle most mobility scenarios. \\
(ii) \textbf{Reacting quickly when blockage happens}: When blockage happens, the goal is to quickly switch
to a receive beam that can receive a reflected beam of the transmit beam.
The mobile simply switches to the previously stored
receive beam that was found to best receive the reflected beam of the current transmit beam.
It continues to work since the BCT of the reflected beam is more than $100$ ms.

\section{BeamSurfer Protocol}\label{beamsurfersection}
\begin{figure}[h]
  \centering
  \includegraphics[width=.8\linewidth,scale=.5]{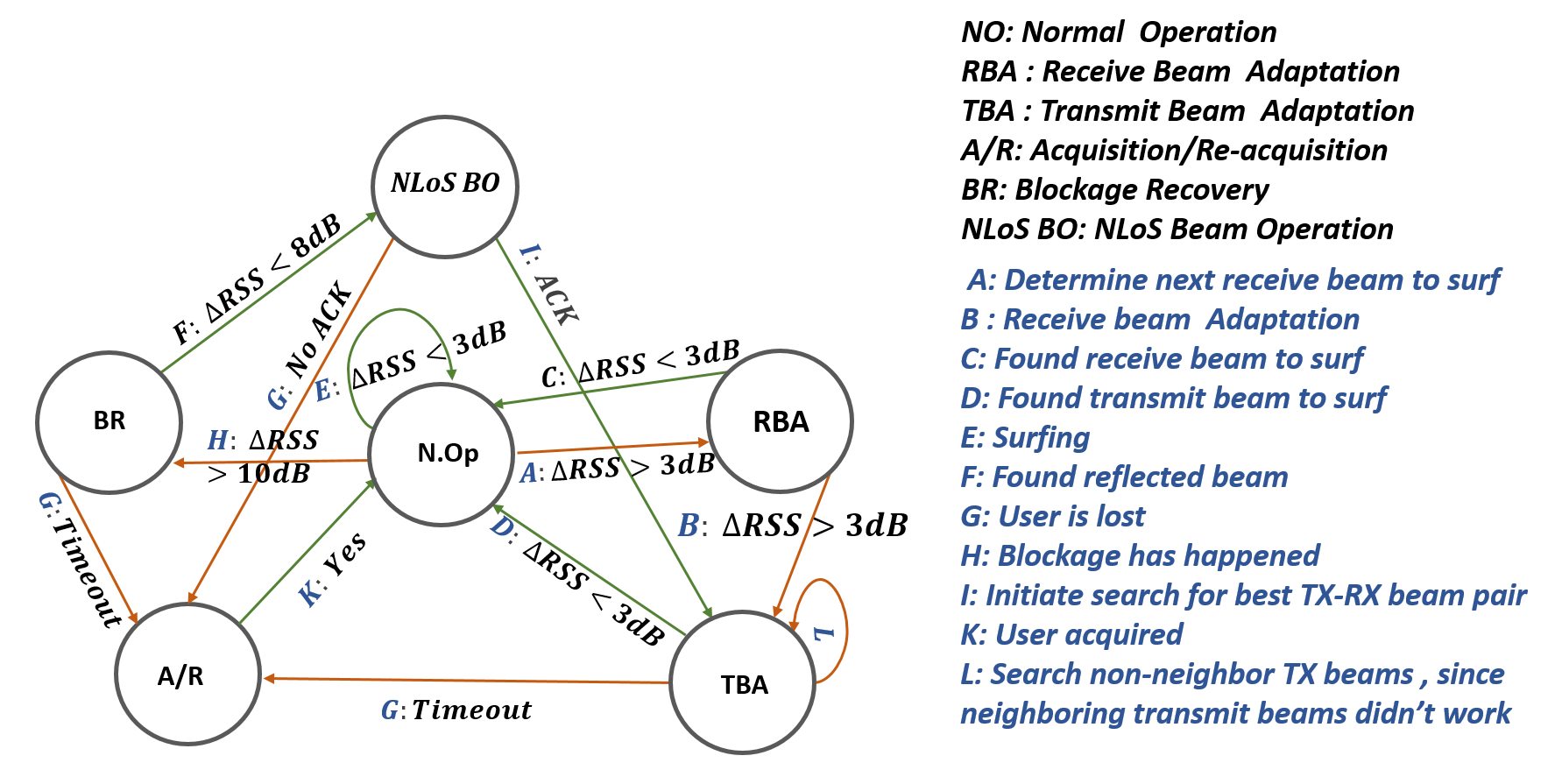}
  \caption{BeamSurfer: State Machine}
  \label{bsflochart}
\end{figure}

\begin{figure}[h]
  \centering
  \includegraphics[width=0.7\linewidth,height=3.4in]{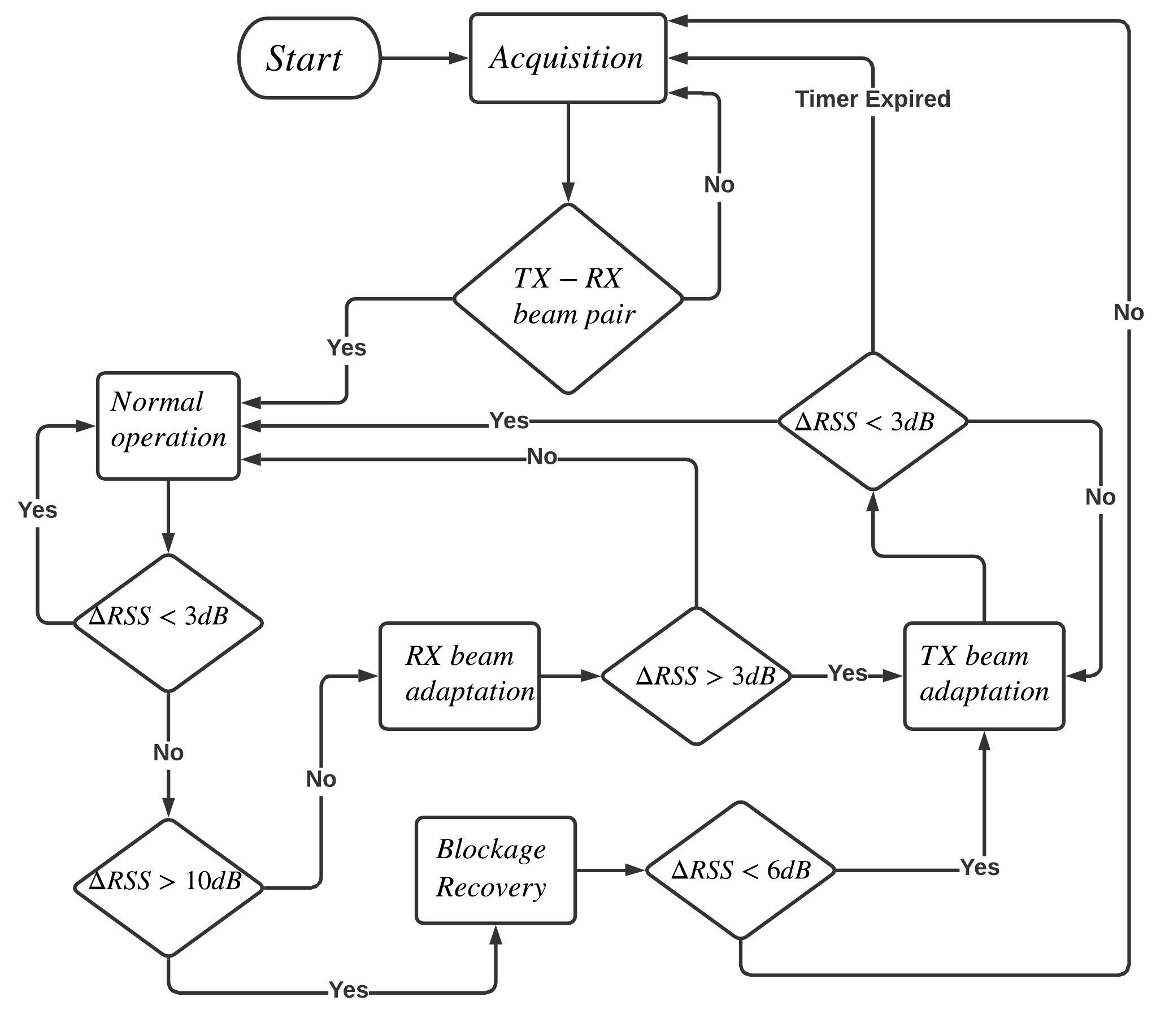}
  \caption{BeamSurfer: Flowchart}
  \label{bsflochart1}
\end{figure}

 We now present the BeamSurfer Protocol for maintaining high received signal strength 
in spite of mobility, and for transient blockages.
 The state machine of the protocol is shown in Fig. \ref{bsflochart}. The flow chart of the protocol is shown in Fig. \ref{bsflochart1}.

N.Op corresponds to ``Normal Operation" using a LoS Transmit Beam from the Base Station,
and an aligned Receive Beam of the user.
Let $\mbox{RSS}_{\mbox{\tiny{Current RB}}}$
denote the signal strength at the time that the current Receive Beam was chosen.
In this state, the mobile samples the signal strength every { $100$ ms}. It continues to use the same receive Beam as long as the signal strength has not dropped by more than $3$ dB { from $\mbox{RSS}_{\mbox{\tiny{Current RB}}}$}. 

If the signal strength of the current receive beam, say beam $k$,  has dropped by more than $3$ dB from $\mbox{RSS}_{\mbox{\tiny{Current RB}}}$, then the protocol moves to RBA: Search for Best Receive Beam. The mobile tests receive beams $k-1$ and $k+1$ for their signal strength, and selects whichever is larger. If that receive beam's signal strength is within $3$ dB of $\mbox{RSS}_{\mbox{\tiny{Current RB}}}$, more precisely if it is greater than  $(\mbox{RSS}_{\mbox{\tiny{Current RB}}}-3)$ dBm, then that receive beam is chosen as the receive beam to use, and the protocol goes to N.Op: Normal Operation.
The value of $\mbox{RSS}_{\mbox{\tiny{Current RB}}}$ is reset to the value of the current received signal strength.

However, if in RBA, both beams $k-1$ and $k+1$ have signal strength less than $(\mbox{RSS}_{\mbox{\tiny{Current RB}}}-3)$ dBm, then the protocol moves to TBA: Search for New Transmit Beam. Let us denote the old Transmit Beam (the one which is to be replaced) by $n$. Upon entering TBA, the mobile asks the base station for transmit beam adaptation. The base station can first test Transmit Beams $n-1$ and $n+1$, as our experiments show that there is high probability of them being the best transmit beam. For each of these Transmit Beams $n-1$ and $n+1$, the mobile first checks both beams $k-1$ and $k+1$ and, if the signal strength doesn't improve, it scans all its receive beams, and chooses the one with the highest received signal strength. 
Suppose that that received signal strength does not constitute a drop of more than $3$ dB from the previous 
$\mbox{RSS}_{\mbox{\tiny{Current RB}}}$, then it is chosen as 
the new Receive Beam and the value of $\mbox{RSS}_{\mbox{\tiny{Current RB}}}$ is reset to current measurement. The protocol moves to N.Op.
On the other hand if even the best receive beam corresponds to a greater than $3$ dB drop from the previous $\mbox{RSS}_{\mbox{\tiny{Current RB}}}$, the mobile continues in this state till the RSS becomes less than $3$ dB. If the signal strength is consistently less than $10$ dB for a timeout period (up to $6$ seconds \cite{_2017_nrrrc}) configured by the base station, then the protocol moves to A/R: Acquisition/Reacquisition.

In A/R, the mobile has been lost, and needs to be re-acquired as though it were a new user. 
This process in 3GPP {5G NR} consists of the Base Station periodically sending out beams in all directions in a predetermined schedule.
For each of these transmit beams, the mobile scans all its receive beams. There is also an opportunity to reply back to the base station by reversing the beams. At the end of the process, the mobile is again reacquired by the base station and a transmit-receive beam pair is chosen.
The protocol moves to N.Op.

If in N.Op, the mobile fails to receive the transmit beam,
due to a large decrease in signal strength, {more than $10$ dB}, then it has suffered blockage.
It then moves to NLoS operation.

In BR, the mobile scans the receive beams to identify an NLoS path. It then stores the
best receive beam for picking up the reflected beam of the current transmit beam. 
If the reflected beam is received on that receive beam with RSS no less than ($\mbox{RSS}_{\mbox{\tiny{Current RB}}}-10$) dBm, then it moves to N.Op.
In NLoS BO, the mobile has already found a path from base station to the mobile that employs reflections. The same path also works in the reverse direction. This path is about $8$ dB lesser than before blockage, but it is still adequate to send control plane packets to the base station.

The mobile sends a control packet to the base station to begin a search for a new transmit-receive beam.  If the base station receives this request, the  base station and user 
continue on to determine a new transmit-receive pair with the best received signal strength. Having a reflected beam hastens the new transmit-receive beam search. In case of mobile blockers, the blockage event is transient and a new beam pair can restore LoS operation. For prolonged blockage, the mobile continues to operate with best transmit and NLoS receive beam pair. {
If either the base station does not receive this request packet or the user does not receive the base station's response for some time, then it concludes that
the user has lost connectivity with base station, and the protocol moves to A/R.
}  Also if the received signal strength is less than ($\mbox{RSS}_{\mbox{\tiny{Current RB}}}-10$) dBm, where ``Current RB" refers to the Receive Beam presently being employed., then the protocol moves to A/R: Acquisition/Re-Acquisition.

In addition to this, in between epochs, the mobile periodically moves to BR and 
scans all receive beams to discover a reflected beam of the current transmit beam that it is using to communicate with the base station,
as a fallback option if there is sudden blockage.
Specifically, BeamSurfer performs a spatial receive beam scan every $100$ ms or every time a transmit beam is changed. In our Testbed it takes $2.5$ ms for the mobile to scan all receive beams and discover the presence of the reflected paths.
It stores the best receive beam for use in case of blockage.

\section{Evaluation of BeamSurfer} \label{Implementation and Evaluation}

\noindent\textbf{A. Performance of Adaptive Beam Alignment of BeamSurfer.}
The two node $60$ GHz software defined radio Testbed  \cite{instruments_2020_introduction} 
is described in Section \ref{testbedsection}. After aligning the transmitter and receiver node beams, we performed mobility experiments.  The receiver node  at 5m distance from transmitter is moved laterally, i.e., perpendicularly,  to the
direction of the transmitter's location at $1.4$ m/s. We dub this mobility scenario as ``Translational motion" below.  We also evaluated BeamSurfer for rotational motion, dubbed ``Rotational motion" below, in which the receiver antenna which is 5 m away from the transmitter is rotated within a $120^\circ$ sector at an angular speed of $120^\circ$/sec. Finally, in the most important scenario, we evaluated BeamSurfer when the receiver antenna is carried by a user employing free hand movements while randomly walking in the coverage sector and 5 m from the transmitter. The random motion with hand movement of the user causes both translation and rotational motion, and allows evaluation of the protocol in a real-world scenario. We dub this mobility scenario as ``Random walk".

\begin{figure*}[t]\label{BeamSurfer's handling of misalignment during user mobility}
	\centering
	\begin{subfigure}[t]{.3\linewidth}
		\centering
		\includegraphics[scale=.55]{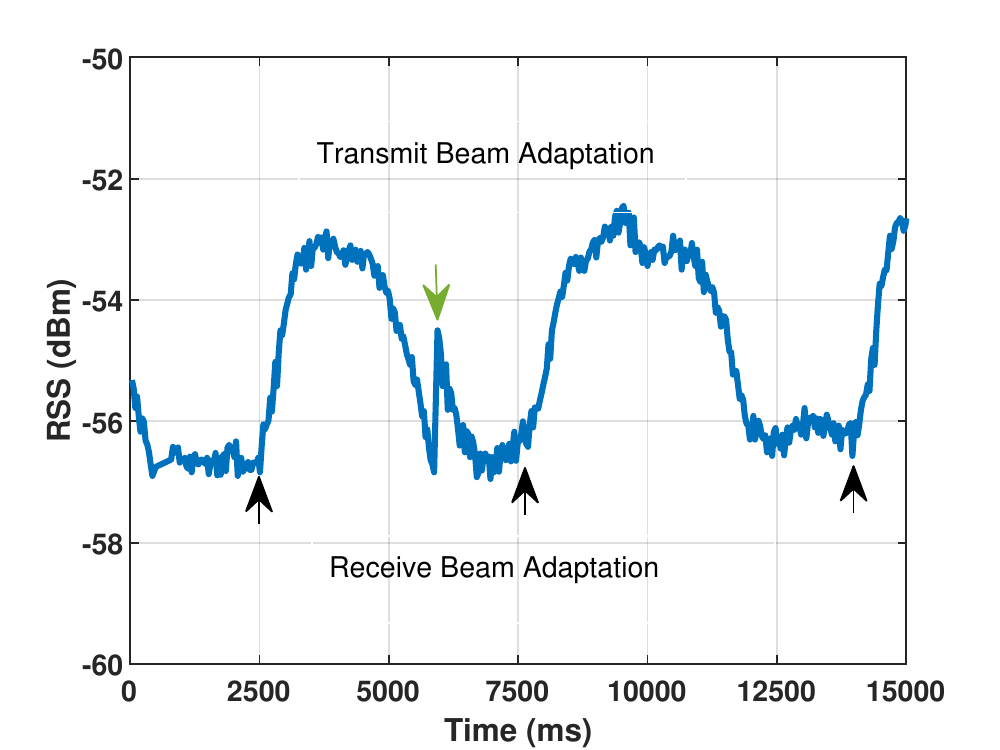}
		\caption{Translational motion with V = 1.4 m/s}\label{fig:1}		
	\end{subfigure}
	\quad
	\begin{subfigure}[t]{.3\linewidth}
		\centering
		\includegraphics[scale=.55]{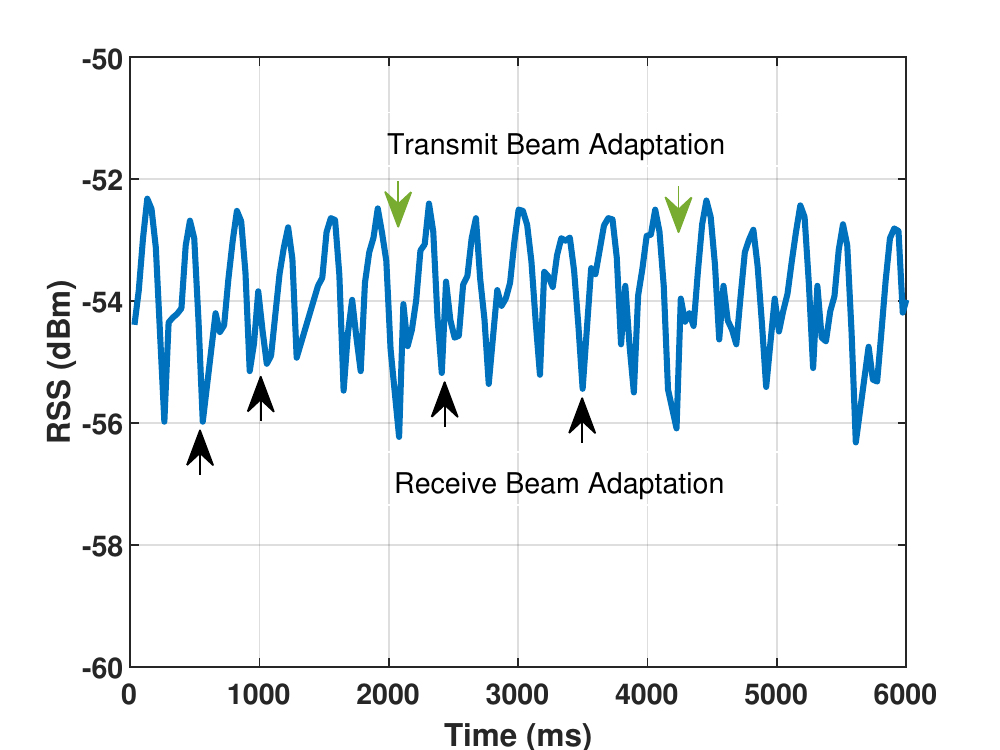}
		\caption{{Rotational motion with $\omega$} = 120 deg/s}\label{fig:2}
	\end{subfigure}
    \quad
	\begin{subfigure}[t]{.3\linewidth}
		\centering
		\includegraphics[scale=.55]{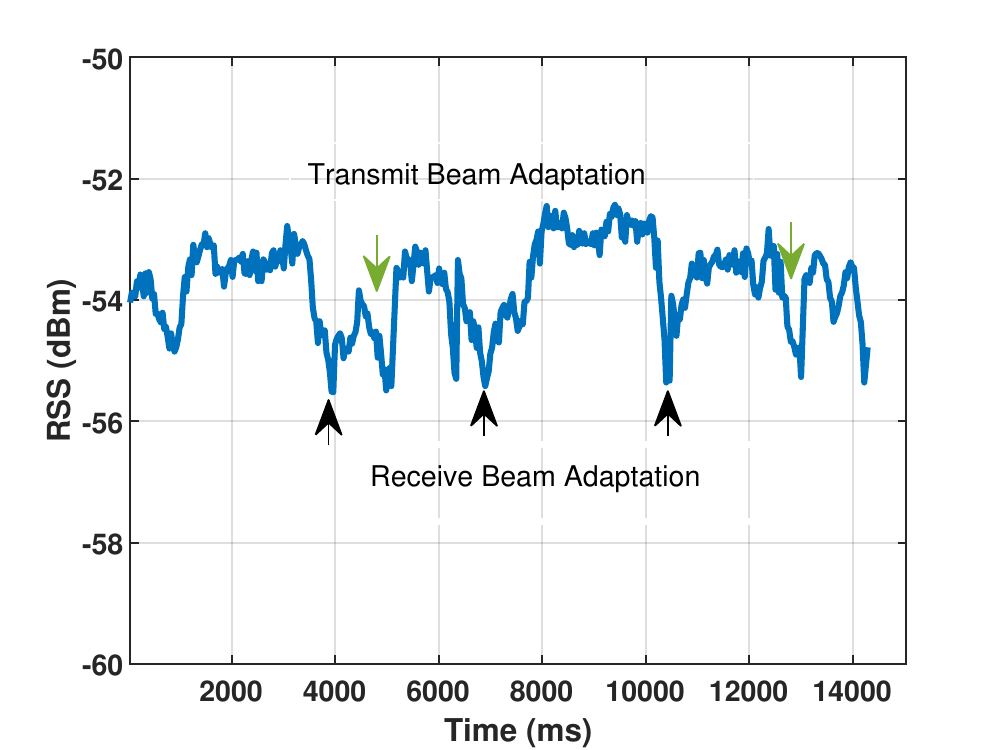}
		\caption{{Random Walk}}\label{fig:3}
	\end{subfigure}
		\caption{BeamSurfer's adaptation to misalignment caused by user mobility}
		\label{beamsurferalignmentresults}
\end{figure*}

Fig. \ref{beamsurferalignmentresults} plots the snapshots of signal strength in one of the experiment trials for each of the three motion patterns. 
We see that BeamSurfer maintains received signal strength variation within about 4 dB of its peak by continually adapting transmit and receive beams. We indicate the time
instants at which BeamSurfer transitions among its states to adapt to mobility.
Both Transmit and Receive Beam adaption instants are marked in the signal snapshots. The receiver first performs receive beam adaptation, switching its beam to one of its neighbor beams, to maintain signal strength. If the receive beam adaptation cannot restore link signal strength, transmitter beam adaptation is triggered. We observe that BeamSurfer's state transitions occur more often during rotational motion, as observed Fig. \ref{beamsurferalignmentresults}b,  in comparison to translational motion. This conforms with our expectation since BCT is smaller for rotational than for translational motion. 
A video demonstration \cite{ganji_10_beamsurfer} of BeamSurfer handling beam adaptation during user mobility is available online \cite{beamsurfer}.  
\captionsetup[subfigure]{labelformat=parens}




\begin{table}[H]
\small 
   \caption{Average deviation of the Received Signal Strength of BeamSurfer Solution from the best beam that an omniscient Oracle would have used.}\label{rssdeviation2}
       \centering

		\begin{tabular}{ |P{8.5 cm}|P{6.5cm}|}
\hline		
\hline
    {Motion } & {Standard deviation (in dB) } \\
 \hline\hline
 Rotational motion at $\omega$ = 60 (deg/s)  & 2.1    \\ 
 \hline
Rotational motion at $\omega$= 120 (deg/s)  & 2.3 \\ 
 \hline

Rotational motion at $\omega$= 240 (deg/s)  & 2.31 \\ 
 \hline

Lateral motion at V = 1.4 (m/s) & 1 \\ 
  \hline

Random walk: User holding phased array walks randomly between transmitter and receiver & 1.21  \\ 
 \hline
\hline
\end{tabular}
\end{table}

In order to evaluate how close to optimal is the behavior of BeamSurfer, we determined the signal strength on the best receive beam at each point along the trajectory of motion, during prior experimentation. This is the signal strength of the beam that an omniscient Oracle would choose. We dub this as ``Oracle's solution." Table \ref{rssdeviation2}
compares the performance of BeamSurfer against this absolute upper bound on performance, for each of the motion scenarios.
It tabulates the average deviation of BeamSurfer's received signal strength from the Oracle's beams,
over 30 trials, for each of the motion pattern. 
It can be seen that
the received signal strength of the beam chosen by BeamSurfer is within $3$ dB of the Oracle's solution $95\%$ of the time. 


\noindent
\textbf{B. Performance of Blockage Recovery by BeamSurfer.}
Keeping the transmitter and receiver nodes $4$ m apart and $2$ m from the wall,  we  placed  the receiver antenna array at $20$ different locations in the layout shown in Fig. 
\ref{layout}.
Then we blocked the LoS paths and checked whether BeamSurfer could receive signals via NLoS paths. To evaluate BeamSurfer's blockage recovery under mobility,  we moved both blockers (person and wooden boards) and receiver array randomly,  $10$ times at each of these locations.

\begin{figure}[h]

	\begin{subfigure}[t]{.5\linewidth}
		\centering
		\includegraphics[width=.9\linewidth]{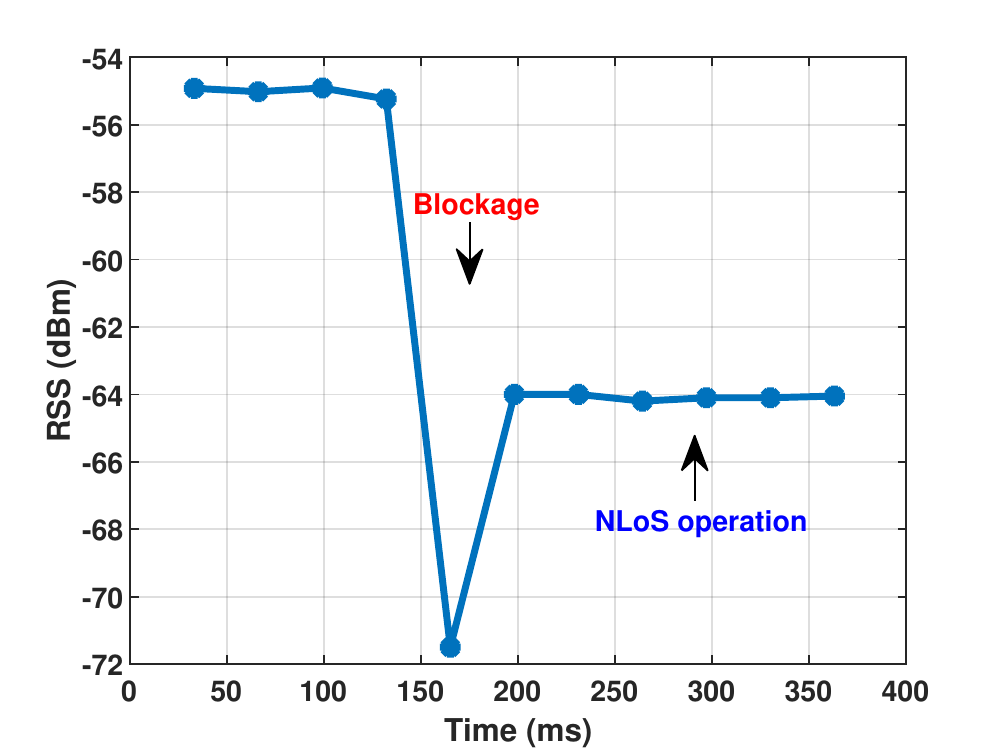}
		\caption{Blockage recovery}\label{fig:2}
	\end{subfigure}
		\quad
	\begin{subfigure}[t]{.5\linewidth}
		\centering
		\includegraphics[width=.9\linewidth]{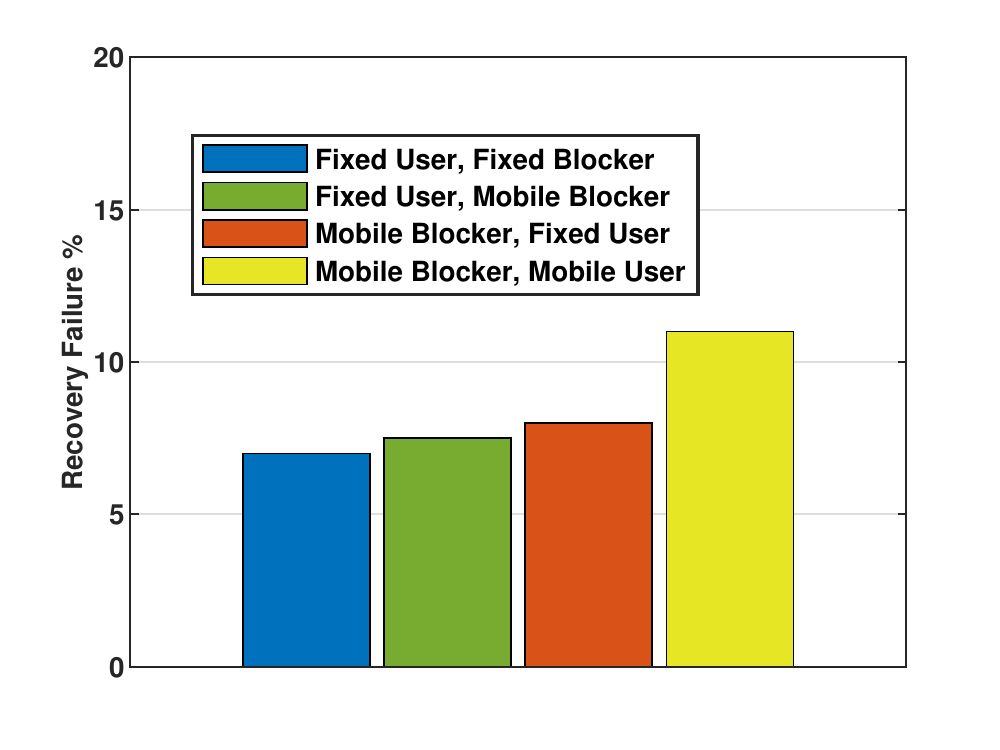}
		\caption{Recovery failure rate}\label{fig:3}
	\end{subfigure}
  \caption{BeamSurfer's blockage recovery}
  \label{blockagerecovery}\end{figure}
We tested BeamSurfer's performance in all the four possible scenarios where either user/blockers can be fixed at certain locations or are mobile.
We measured  the signal strength at those locations, and confirmed that BeamSurfer was in fact retrieving the receive beam from memory. Fig. \ref{blockagerecovery}a shows 
the received signal strength during the blockage for a segment in the experiment. The state transitions are annotated, and show that BeamSurfer recovers its signal strength by switching to a NLoS path. Despite being a simple approach, BeamSurfer's blockage recovery failure rate as shown in Fig. \ref{blockagerecovery}b is under $11.5\%$, outperforming existing works \cite{beamspy} which utilise a large memory to store path parameters and compute power to make predictions to recover from blockage. {For the case of a static user and a static blocker, BeamSpy \cite{beamspy} is reported to recover link during blockage $60\%$ of the time, apparently in static experiments, whereas BeamSurfer recovers $93\%$ of time in experiments under user mobility.}

\begin{figure}[h]

	\begin{subfigure}[t]{.5\linewidth}
		\centering
		\includegraphics[width=.9\linewidth]{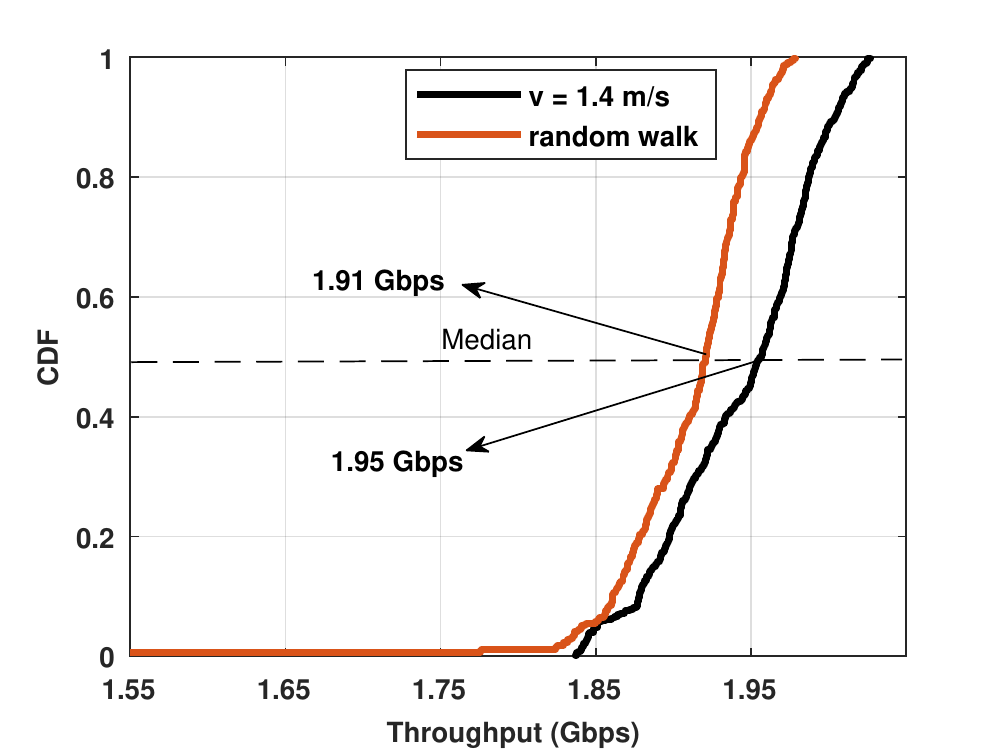}
		\caption{Translational Motion}\label{fig:2}
	\end{subfigure}
		\quad
	\begin{subfigure}[t]{.5\linewidth}
		\centering
		\includegraphics[width=.9\linewidth]{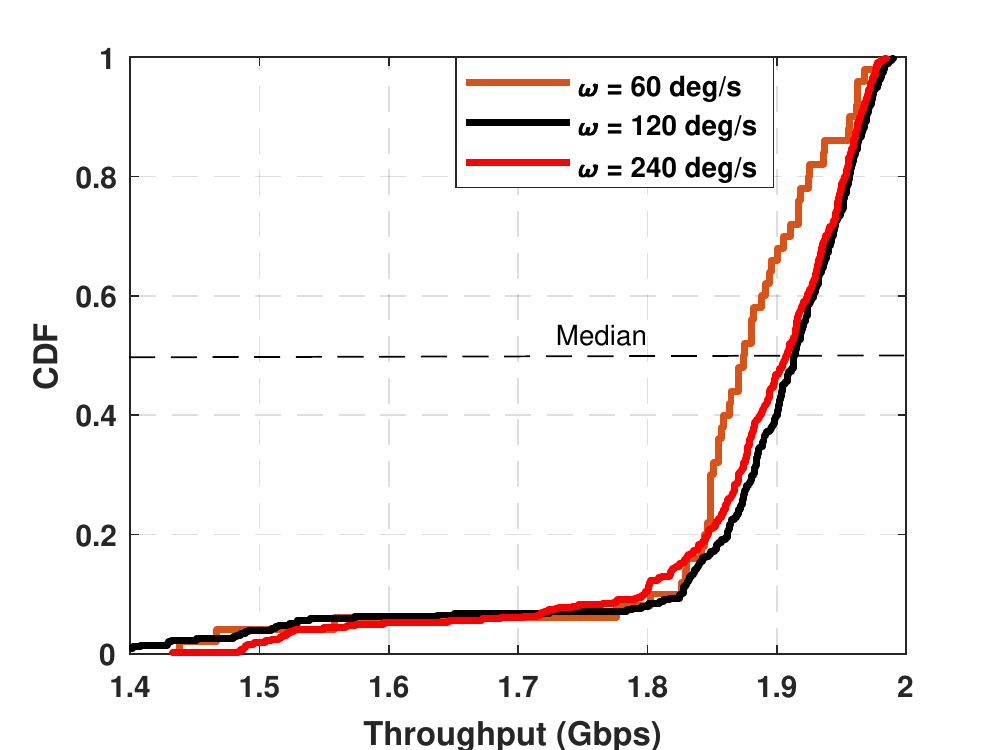}
		\caption{Rotational Motion}\label{fig:3}
	\end{subfigure}
	\caption{The Cumulative Distribution Function (CDF) of the Throughput for Users with Translational and Rotational Motion. { The median throughput for Translational motion at 1.4m/s is 1.95 Gbps, and that for Random walk is 1.91 Gbps.}}\label{Throughput}
\end{figure}

\noindent
\textbf{C. System Level Evaluation.}
To perform system level evaluation, we sent turbo encoded randomly generated bits with $3/4$ rate and $16$ QAM MCS from the transmitter node to the receiver node. We repeated the above translational and rotational motion experiments within the same layout in both indoor and outdoor environments.  We compared the throughput after enabling BeamSurfer at the receiver with the Oracle's solution of $2$ Gbps, i.e., the best aligned transmitter and receiver beams.  
{
Fig. \ref{Throughput}a presents the CDF of the downlink throughput during human walk with speed of V = $1.4$ m/s. We present two scenarios, first, while a user is walking in a linear  trajectory perpendicular to the LoS path between transmitter and receiver, and second, while walking randomly between transmitter and receiver. BeamSurfer provides median throughput of $1.95$ Gbps and $1.91$ Gbps, respectively, during translational walk (V = 1.4 m$/$s) and random walk. Fig. \ref{Throughput}b shows the downlink throughput when the receiver is rotated within a $120^\circ$ sector with angular speeds of 60, 120 and 240 $deg/s$. We can observe that BeamSurfer obtains at least $1.4$ Gbps,  and that its median performance is within $92.5\%$ of Oracle throughput in all the scenarios.
 Also, the throughput is within $150$ Mbps of the { Oracle's upper bound} of 2 Gbps on the maximum achievable throughput, $90\%$ of the time under
translational motion, and 80\% of the time under
rotational motion. } 

\noindent
\textbf{D. Performance Comparison.} The first issue for comparison of beam alignment algorithms is whether they require omni-directional reception of received signal strength measurements, or can work with directional receivers.  
%
Schemes that require omni-directional measurements have much lower link signal strength and therefore have significantly reduced range.
To determine the LoS beam between transmitter and receiver,  Swift link \cite{Newref1} and FALP \cite{Newref6} require signal measurements using an omni-directional receiver while transmit beam patterns are quasi omni-directional. Quasi omni-directional transmit beam patterns have very low directional gain in any particular direction.   Hierarchical Beam Alignment (HBA) \cite{Newref2} starts receiver measurements using omni-directional patterns and gradually narrows the receive beam. Agile Link \cite{beamalign1} uses specially designed beam patterns with lobes in multiple directions, whereas both BeamSurfer and plain Exhaustive Search, employ highly directional beams with a single main lobe  for each measurement, thus allowing operation at greater range. The SNR and therefore data rate using both Exhaustive Search and BeamSurfer are always higher by a margin of 20 dB  than Swift link \cite{Newref1} and FALP \cite{Newref6} as they require an omni-directional receiver. The receive mode in IEEE 802.11  ad. significantly reduced network range and the  standard itself was quickly superseded by IEEE 802.11 ay which advocates directional antenna use. It is also important to note that there is no such explicit omni-directional mode in 5G New Radio.  
Some shortcomings  of omni-directional receive mode of 802.11 ad are presented in \cite{shortcomings}.

Next, we compare the number of measurements required for each scheme.
We compare them  using the same system model as described in Swift Link \cite{Newref1} and FALP \cite{Newref6}, a Uniform Planar Array of 32x32 at Transmitter, at a distance 10 m away from a transmitter  operating at 28 GHz, where the total number of beam patterns for the array are 1024. 
 Table \ref{num_measurements} presents the maximum number of measurements required by various works in the literature to re-align or identify a beam when a receiver beam is misaligned with that of the transmitter beam, i.e., when the user moves out by just one beam width in azimuth.
Compressive sensing based methods are well suited for initial beam search, but for beam adaptation during mobility, which BeamSurfer is addressing, they seem to require large number of measurements. Initial search can be longer, but once the mobile is connected to the base station, beam adaptation must be quick as measurements need network resources and delayed beam alignment reduces the throughput of mobile. 
 Beam-forecast \cite{beamalign2}, first creates a geometric model of environment within a certain distance from an anchor location. Based on the model, it identifies a set of potential beam-pairs at any given neighbor location within 3 m of the anchor point to narrow down the number of beams to be searched. The highest RSS beam-forcast \cite{beamalign2} can achieve even after using more than 100 beam-pairs is 1.9 dB below the Oracle RSS due to the discrepancy between the model and real environment.

\begin{table}[h]
\small 
   \caption{Maximum number of measurements to achieve beam alignment}\label{num_measurements}
\begin{center}
\begin{tabular}{ | l | c |  }
\hline
\hline
State of the art & Maximum Number of Measurements  \\ \hline \hline
 BeamSurfer & 8  \\  
 Beam-forecast & 40\\
 HBA & 63 \\
 Swift Link & 70 \\
 FALP & 70 \\
 
 Agile Link & 110 \\
 Exhaustive Search & 1024 \\
 \hline
\end{tabular}
\end{center}

\end{table}

\noindent
\textbf{E. Hardware Implementation Performance Comparison.}
To adapt the receiver beam,  FALP \cite{Newref6} and Swift link \cite{Newref1} need several measurements after which an optimization solver is used to identify an aligned beam.
Using the source codes available for Swift Link \cite{Newref1}, FALP \cite{Newref6},  computation times for the respective optimization solvers which output the beam decision on a PC with Intel i7 processor running at 4.2 GHz clock rate and 16 GB RAM are 10.5 and 5 seconds, respectively. 
In contrast, BeamSurfer only requires received signal strength and is completely implemented on FPGA.  

\noindent
\textbf{F. Mobility and Experimental Validation Performance.}
       BeamSurfer is implemented on FPGA, and performance has been        demonstrated during actual mobility experiments. 
       Swift Link \cite{Newref1}, FALP \cite{Newref6} and HBA \cite{Newref2} do not present any  experimental validation. They also do not present any simulation results for performance under mobility.
\section{Testbed and Implementation}\label{testbedsection}
\begin{figure}[h]

	\begin{subfigure}[t]{.5\linewidth}
		\centering
		\includegraphics[width=.8\linewidth,keepaspectratio]{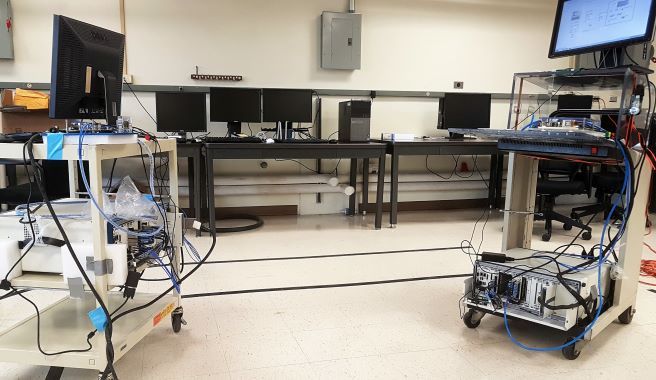}
		\caption{Transceivers}\label{t1}
	\end{subfigure}
		\quad
	\begin{subfigure}[t]{.5\linewidth}
		\centering
		\includegraphics[width=.4\linewidth,height=1.55in]{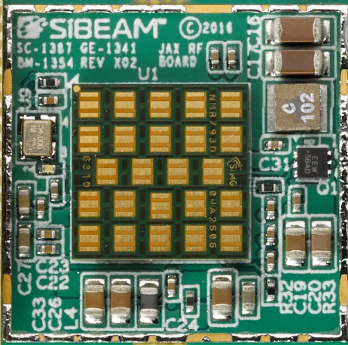}
		\caption{Sibeam Array}\label{t2}
	\end{subfigure}
	\caption{Testbed and Phased Array}\label{Transceivers}
\end{figure}

Our Testbed is a National Instruments software defined mm-wave transceiver system with two nodes. 
\begin{figure}[h]
  \centering
  \includegraphics[width=.8\linewidth,scale=.4]{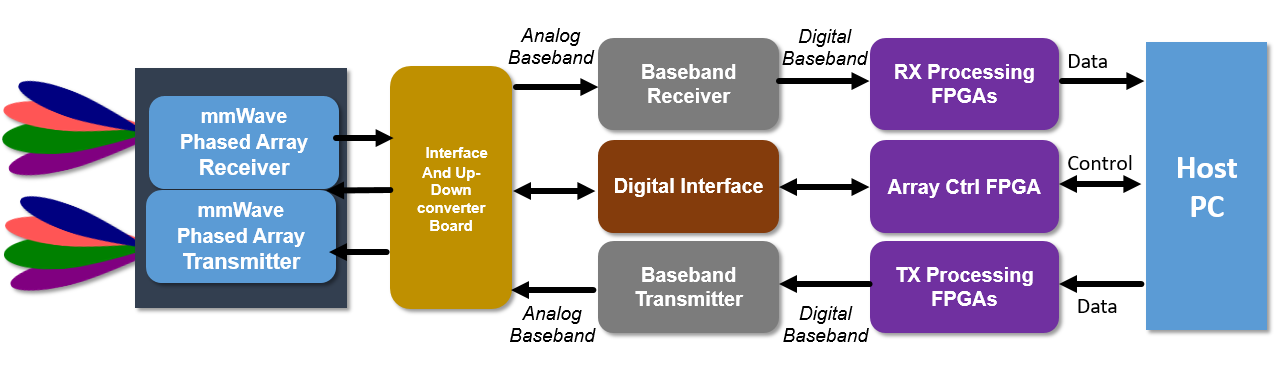}
  \caption{Testbed Functional Diagram}
  \label{testbed1}
\end{figure}
Fig. \ref{testbed1} presents the functional diagram of the testbed.  The node is built using a chassis with high speed backplane interconnections. A host computer with an Intel processor interacts with multiple FPGA cards that can communicate with each other. Each transceiver node is connected with a SiBeam phased array. The SiBeam phased array shown in Fig. \ref{t2}  operating at $60$ GHz has $24$ antenna elements; $12$ each for TX
and RX. Wide and narrow beams can be formed by changing the number of active antenna elements. To produce narrow beams we use upper $10$ elements forming a uniform planar array. A series of $4$ elements are used to form wide beams. For our experiments, we used azimuth beam-forming ``codebook" with $25$ equally spaced wide or narrow beams, roughly covering a sector of $120$ degrees, obtained by programming  the $2$ bit phase weights.  


For making fine grain measurements, we transmit a single carrier waveform in a slot that has a duration of $100$ microseconds. A frame is created with $100$ slots, with the first $50$ slots for downlink and rest for the uplink operation. Four slots within the $50$ are used to time synchronize base station and user nodes.  We send a Zadoff Chu sequence known to the receiver in these 4 slots to acquire timing of the frame and thereby slots.
We implemented BeamSurfer in Labview at the user node in our Testbed. For the transmit beam adaption,
we created a small control information packet in Labview on the node running BeamSurfer. The base station node upon receiving this packet either initiates or terminates the transmit beam sweep.
\section{BeamSurfer application in 5G Newradio} \label{BeamSurfer application in 5G Newradio} 
Beam Management at the base station, called gNodeB, and user equipment (UE) in $5$G Newradio standards involves a beam search at the UE for initial beam alignment with the gNodeB, and continuous beam tracking both at the gNodeB and the UE to maintain beam alignment to counter user mobility.
The $5$G NR standards provide a framework and opportunities both for gNodeB and UE to identify an aligned beam pair and further maintain the alignment. If the gNodeB or the UE fails to maintain an aligned beam pair, then communication gets disrupted and the UE must start again with initial beam search.
 

In every beam, gNodeB transmits synchronization signal blocks (SSB) for timing synchronization, cell related broadcast information and all the control plane information related to scheduling. The UEs first sweep their beams to discover a gNodeB beam and acquire the timing and control information \cite{_2017_nr2}. The gNodeB uses up to $64$ different beams within a sector.  To accomodate a large number of UEs, communication between gNodeB and UEs happens within a $10$ ms duration frame.  A frame is partitioned into slots. Within these slots uplink/downlink or control communication takes place via OFDM symbols.  All the SSB beams are transmitted within one half of a frame with either two or four SSBs in a slot, and this pattern is repeated every two frames i.e., $20$ ms.   Based on the deployment configuration in mm-wave/FR2 bands \cite{bs_fr2}, the number of slots in a frame can be either $80$ or $160$.
 The current 5G standards in mm-wave bands are time-duplexed, meaning the bidirectional communication between gNodeB and the UE happens one after the other but not at the same time. 
 
The UE during initial acquisition searches for the gNodeB beams by synchronizing with SSB timing. Initially, the locations of these SSBs in a frame  are unknown, and UE has to listen on a receive beam for at least $20$ ms. To identify all the $64$ SSB/gNode beams, it takes $1.28$ seconds.
 However,  gNodeB can change the periodicity of the SSB beams to $5$ ms from $20$ ms after a UE is connected it. A/R, Acquisition/Reacquisition, in BeamSurfer alignment handling performs initial search and time synchronizes the UE with gNodeB. 

\noindent
{\textbf{A. Beam Management in 5G Newradio.}}
After initial Beam alignment during A/R state, both the gNodeB and UE must maintain aligned beams to avoid link outage. To accomplish that, the gNodeB schedules a set of beams with with SSB or UE specific Channel State Information reference signals (CSI-RS). UE measures various signal statistics such as received signal strength and SNR for these beams and reports them back to the gNodeB. 
While the measurement reports help the gNodeB to manage its beam, UE can decide its beams while making the measurements. UE is free to change its beams as and when necessary to track the pre-assigned gNodeB beam. RBA in BeamSurfer protocol is most used in scenarios involving user mobility, hand movements and head rotation, which quickly result in beam misalignment at UE. During RBA, whenever UE detects a $3$ dB drop in the signal strength, it immediately switches beams to either of the neighbor beams. It then holds on the best neighbor beam until a further drop in signal strength is detected. In case the received signal strength is not restored in RBA state, as in the case of lateral translation user motion, the protocol state moves to TBA in which gNodeB switches transmit beams based on the channel quality reporting.
 
As the user moves, the beams might get blocked, by the user's own body or obstacles in the environment. To avoid link outage in case of blockage, the UE moves to BR state every $100$ ms to harvest reflected radiation. With knowledge of SSB and CSI-RS beam schedules, unlike in A/R state, the UE performs a spatial scan to identify reflected beams. It stores the best such reflected beam in the memory to employ when blockage happens. 

When there is no anomaly in link signal strength, the UE continues to operate in N.Op state.
\section{Concluding Remarks}
In the millimeter-wave band, the narrow transmit and receive beams get misaligned by translational or rotational mobility, and can get blocked by a person or object.
Disrupted communication between the user and base station may cause service disconnection,  requiring a scan over all transmit-receive beam pairs for the user to reconnect to the base station, a relatively long process,
and disrupting applications. 

BeamSurfer's beam management protocol is based on experimental measurements of how received signal strength changes with mobility, beam direction, and on reflected paths.
It is targeted to the use case of users walking near an indoor base station, though its misalignment mechanism can also be used outdoors..
It is a lightweight protocol that only uses information that is 
normally obtained during communication, without requiring any additional sensors such as for motion.
It handles mobility by continually making small one-step adjustments to the receiver
and transmitter beams, employing the peak gains of
main lobes of the aligned beams.
Evaluations show that its beam alignment protocol 
successfully chooses transmit-receive beams that are at
all times within $3$ dB of what an omniscient Oracle would choose,
in both indoor and outdoor environments.
To handle blockage in indoor environments, BeamSurfer pre-chooses a good reflected beam to use if there is a blockage. The reflected beam can coordinate base station and user to recover from blockage.
 Experiments show that such reflected beams with sufficient received signal strength to sustain control plane communication are typically present for any transmit beam.
The LoS beam alignment protocol of BeamSurfer can also
be used outdoors, as verified experimentally in this work.

In outdoor environments, however, there are no reflected beams, so
blockage recovery with the same base station is not possible.
More generally, whether indoors or outdoors, when the same base station can no longer be used to maintain a connection, there will
need to be a handoff. Such millimeter-wave handoff is a topic of current
research.\newline\\

\bibliographystyle{IEEEtran}
\bibliography{IEEEabrv,references}
\end{document}